\documentclass[conference]{IEEEtran}
\IEEEoverridecommandlockouts
\usepackage[table,svgnames,dvipsnames]{xcolor}

\usepackage{cite}
\usepackage{amsmath,amssymb,amsfonts}
\usepackage{algorithmic}
\usepackage{graphicx}
\usepackage{textcomp}
\usepackage{xcolor}
\usepackage{tikz}
\def\BibTeX{{\rm B\kern-.05em{\sc i\kern-.025em b}\kern-.08em
    T\kern-.1667em\lower.7ex\hbox{E}\kern-.125emX}}

\graphicspath{{figs/}{figures/}{pictures/}{images/}{./}} 

\usepackage{tabu}                      
\usepackage{booktabs}                  
\usepackage{lipsum}                    
\usepackage{mwe}                       

\usepackage{adjustbox}
\usepackage{array}

\usepackage{mathptmx}                  
\usepackage{xspace}
\usepackage{soul}

\usepackage{hyperref}

\usepackage{makecell}

\usepackage{soul}

\newcommand{\hlc}[2][yellow]{{%
                  \colorlet{foo}{#1}%
                  \sethlcolor{foo}\hl{#2}}%
}
\newcommand\qtc[1]{\hlc[SeaGreen!30]{``#1''}}
\newcommand\qtu[1]{\hlc[ProcessBlue!10]{``#1''}}
\newcommand{\cxx}[1]{\textbf{C$_{\scriptstyle\textit{#1}}$}}
\newcommand{\uxx}[1]{\textbf{U$_{\scriptstyle\textit{#1}}$}}

\newcommand{\parahead}[1]{\paraheadd{#1}.}
\newcommand{\paraheadd}[1]{%
    \vspace{0.5em}%
    \noindent%
    \textbf{\textit{#1}}%
}

\usepackage[normalem]{ulem}
\definecolor{linkColor}{HTML}{438C57}
\setuldepth{Berlin}
\newcommand\asLink[2]{\textcolor{linkColor}{\href{#1}{\ul{#2}}}}

\usepackage{tcolorbox}
\usepackage{multirow}

\definecolor{synthesisColor}{HTML}{f0f0f0}

\tcbuselibrary{breakable}
\tcbset{
  width=0.98\columnwidth,
  halign=justify,
  center,
  breakable,
  colback=synthesisColor    
}

\newcommand{\revised}[1]{#1}

\newcommand{\numCreators}{11}
\newcommand{\numUsers}{9}

\makeatletter
\DeclareRobustCommand\onedot{\futurelet\@let@token\@onedot}
\def\@onedot{\ifx\@let@token.\else.\null\fi\xspace}

\def\eg{{e.g}\onedot} 
\def\ie{\emph{i.e}\onedot} 
 
\def\etc{{etc}\onedot} 
 
\def\etal{{et al}\onedot}
\def\etals{{et al}\onedot's}
\newcommand{\ala}{\`a la}
\makeatother

\newcommand{\toolName}[1]{\textsc{#1}}
\newcommand{\plotly}{\toolName{Plot.ly}}
\newcommand{\altair}{\toolName{Altair}}
\newcommand{\vl}{\toolName{Vega-Lite}}

\begin{document}
\renewcommand{\sectionautorefname}{Sec.}
\renewcommand{\figureautorefname}{Fig.}
\let\subsectionautorefname\sectionautorefname
\let\subsubsectionautorefname\sectionautorefname

\title{Considering Visualization Example Galleries}

\author{\IEEEauthorblockN{Junran Yang}
\IEEEauthorblockA{
\textit{University of Washington}\\
Seattle, USA \\
junran@cs.washington.edu}
\and
\IEEEauthorblockN{Andrew M. McNutt}
\IEEEauthorblockA{
\textit{University of Washington}\\
Seattle, USA \\
amcnutt@cs.washington.edu}
\and
\IEEEauthorblockN{Leilani Battle}
\IEEEauthorblockA{
\textit{University of Washington}\\
Seattle, USA \\
leibatt@cs.washington.edu}
}

\maketitle

\begin{abstract}
Example galleries are often used to teach, document, and advertise visually-focused domain-specific languages and libraries, such as those producing visualizations, diagrams, or webpages.
Despite their ubiquity, there is no consensus on the role of ``example galleries'', let alone what the best practices might be for their creation or curation.
To understand gallery meaning and usage, we interviewed the creators (N=\numCreators{}) and users (N=\numUsers{}) of prominent visualization-adjacent tools.
From these interviews we synthesized strategies and challenges for gallery curation and management (\eg\ weighing the costs/benefits of adding new examples and trade-offs in richness vs ease of use), highlighted the differences between planned and actual gallery usage (\eg{} opportunistic reuse vs search-engine optimization), and reflected on parts of the gallery design space not explored (\eg\ highlighting the potential of tool assistance).
We found that galleries are multi-faceted structures whose form and content are motivated to accommodate different usages---ranging from marketing material to test suite to extended documentation.  
This work offers a foundation for future support tools by characterizing gallery design and management, as well as by highlighting challenges and opportunities in the space (such as how more diverse galleries make reuse tasks simpler, but complicate upkeep).

\end{abstract}


\section{Introduction}

Examples are a powerful medium of knowledge sharing in both design and coding. Designers and artists look into examples as sources of inspiration~\cite{bako_understanding_2022, lee_designing_2010, xu_ideaterelate_2021} and as tools to communicate concrete feedback~\cite{kang_paragon_2018}. In software engineering, code examples provide resources to understand how things work~\cite{nasehi_what_2012}, learn usage patterns~\cite{thayer_theory_2021, battle2022exploring}, and as a means of researching opportunistic reuse~\cite{brandt2009two} (such as how developers will copy and tweak examples from StackOverflow).
Leveraging the power of examples, \emph{galleries} are a common entry point into many domain-specific languages and libraries---we use the latter of these terms throughout for simplicity.
Example galleries are especially common for libraries which include a visual component, such as those that produce visualizations (such as Vega~\cite{satyanarayan2015reactive}), diagrams (like Penrose~\cite{ye2020penrose}),  art (\ala{} Processing~\cite{reas2007processing}), and even webpages (\eg{} Idyll~\cite{conlen2018idyll}). 
Compared to individual examples, galleries
illustrate the gamut of capabilities
through curated collections of code with accompanying visual outputs.



Galleries come in many forms and shapes.
Some consist of every conceivable usage, as with \toolName{Plot.ly}.  Others are tightly managed sets of minimal examples that are meant to advertise the tool, as with ~\toolName{Deck.gl}.
In some contexts~\cite{bako_understanding_2022}, galleries act as tutorials or starter code. 
In others, galleries are intertwined with the library infrastructure (\eg{} as test suite). 
These discordant approaches make it hard to identify \textbf{what makes an example gallery effective}.
Questions like \emph{how many examples should be included?} and \emph{which examples are necessary?} remain unanswered.
The interaction between ``official'' channels and ``folk'' documentation (\eg{} StackOverflow) are unclear. 
Without clear guidelines creators are left to copy the works of other gallery creators, who may have different motivations and resources than they do.



We explore these questions via an interview study. This consisted of semi-structured interviews with (N=\numCreators{}) authors of galleries as well as (N=\numUsers{}) \emph{users} of a subset of those example galleries.
We identify three central reasons why galleries are formed: to advertise the library, to accelerate library usage via reuse, and to support gallery sustainability by acting as a test suite. 
We then analyze a series of common cross-cutting issues that appear that are affected by these various curatorial stances including the contents of the gallery, how the gallery is maintained over time, and role of education and communication in gallery architectural and organization.
Across these analyses we observe strategies and challenges for gallery construction and management (such as the burdens of maintaining the examples during library development), highlight the differences between planned and actual gallery usage (such as between search engine optimization and opportunistic reuse~\cite{brandt2009two}), and highlight opportunities in the space of example galleries (such as providing integrated generative assistance alongside gallery search).  
Broadly, the effectiveness of a gallery seems to depend on the goals of both the creators (\emph{do they integrate examples into the test suite or just the documentation?}) and the user community around it (\emph{how do they tend to use/contribute examples for this library?}). 

We identify practices that make it easier to form and maintain galleries in the future (summarized in \autoref{fig:summary}). For instance, diversifying the types of tasks covered in a gallery can make it easier for users to find specific tasks that are relevant to them and provide a more useful test cases for creators. Yet, smaller galleries are easier to maintain and put less burden on the developers. 
Like most socio-technical artifacts, there are few unambiguous guidelines; for a system with fixed resources, compromises will need to be made to support conflicting usage patterns (\eg{} maintainers vs users). 
Our work in this paper is to make the space of trade-offs clearer, so that design decisions might be made more clearly.

\section{Background and Related Work}
\label{sec:background}

Here we set definitions and review prior work in this space.



\parahead{Documentation Formats} Galleries are but one part of a constellation of different types and forms of documentation written by software creators and consumed by end users~\cite{geiger_types_2018}. 
These include various common forms which describe here to provide language for comparing and contrasting with galleries.
Diataxis~\cite{ProcidaDiataxis} taxonomizes documentation types based on task and form.
We draw on this theory in organizing our analysis of galleries as documentation form.
We use examples from \toolName{Vega} as exemplars of these forms.


\emph{Tutorials and usage documentation} demonstrate step-by-step processes someone can follow to achieve goals specified by the software authors such as installing the DSL or recreating a recommended example. 
They are a common entry point to documentation, designed for beginners to learn by doing.
Code examples in tutorials focus on capturing the process. They are organized into digestible chunks and often display the intermediate results of the segmented code~\cite{head_composing_2020}. Textual information and visual output are used to help elicit the context of changes.  
For instance, Vega's \asLink{https://vega.github.io/vega/tutorials/bar-chart/}{``Let's Make A Bar Chart Tutorial''}~\cite{vega-bar-chart} is an example of this usage.
While tutorials highlight step-wise changes and reference code examples focus on specific usage and assumed basic understanding of the library, gallery examples are self-contained so that users can copy and paste the code to successfully execute it. 
Some of gallery examples focus on a single learning objective, while others (\ie showcases) include more narrativized context.

\emph{References} describe the function of features exposed by the library such as functions, methods, attributes and parameters.
This most basic form of documentation characterizes the specific way in which the user can interact with the library aspect by aspect.
Libraries like Sphinx~\cite{sphinx} automatically generate documentation from code comments.
Vega's \asLink{https://vega.github.io/vega/docs/}{``Documentation''}~\cite{vega-docs} section demonstrates this pattern. 
In contrast to galleries, which feature standalone examples, references are short and focused on an aspect of a particular feature. 


\emph{Explanation and discussion} elucidate a topic in a blogpost style. Unlike a tutorial, the goal is not to accomplish a specific task but to broaden the understanding of domain knowledge from a high level or a different perspective~\cite{ProcidaDiataxis}. A library architecture overview or a background section would fall into this category, such as in 
\asLink{https://observablehq.com/@vega/how-vega-works}{``How Vega Works''}~\cite{vega-internal}. 
It describes the way in which the internals of the library operates. While informative, these resources do not visually demonstrate functionality the way that galleries do, nor do they offer the re-usable cookbook-like guide~\cite{ProcidaDiataxis} to accomplish tasks.

\parahead{Documentation Studies}
Code examples and documentations are essential for learning programming and accomplishing programming tasks.  
Nasehi \etal{}~\cite{nasehi_what_2012} find that well-received answers on StackOverflow often incorporate more code examples with explanation.
Links to existing resources that provide reliable background knowledge are frequently shared, indicating that the general audience have difficulties in finding useful information. 
Nam \etal~\cite{nam24UnderstandingLogs} analyze documentation page-view logs from popular cloud-based services, finding a correlation between visit patterns and user characteristics. 
Other work investigates the challenges using code examples for different user groups, including lack of API knowledge~\cite{thayer_theory_2021}; identifying relevant parts, integrating and adapting existing code ~\cite{ichinco_exploring_2015, wang_novices_2021}. 
While code examples are a key part of galleries, we investigate the interaction with example collections as a whole.




Studies have explored how documentation might be enhanced\revised{ by supporting sustainable maintenance~\cite{horvath2023support}, annotation~\cite{horvath2022understanding}, automated creation~\cite{head2015tutorons, le_moulec_automatic_2018, crichton2020documentation}, enhancements to the development environment~\cite{potter2022contextualized, mcnutt_study_2023}, among others}. 
For instance, ActiveDocumentation~\cite{mehrpour_active_2019} compares documented design rules against users' codebase to provide feedback. 
Glassman \etal~\cite{glassman_visualizing_2018} develop a visualization for summarizing code collections in a synthetic code skeleton, which can help users explore how others have used an unfamiliar API. 
Jernigan \etal~\cite{jernigan_principled_2015} provided debugging hints to end-users who are not interested in programming. 
Our work explore effective galleries as the premise towards dynamic enhancement. 
Related to some of the tasks galleries attempt to serve are code search tools (\eg{} Blueprint~\cite{brandt2010example} or Calcite~\cite{mooty2010calcite}).
Whereas search supports identification of relevant examples, galleries support browsing what is possible within a given context.  


Several works studied the design of gallery-like entities. 
Bako \etal ~\cite{bako_understanding_2022} explore how visualization designers search and use examples. They find people select examples by examining the trade-off between the effectiveness and aesthetics, and reuse them by modifying and merging the visual elements (\ala{} opportunistic programming~\cite{brandt2009two}). 
Kruchten \etal{}~\cite{kruchten_metrics-based_2023} introduce a specialized form of gallery for comparing features among visualization libraries.
We consider the more common case of single-tool multipurpose galleries.
\section{Methodology}
\label{sec:study-design}

We next describe the methods used in our studies.

\parahead{Gallery Survey}
To situate our interview study with users and creators of example galleries, we develop a baseline answer to the question \emph{what is a gallery?}
by surveying 14 visualization-adjacent notations (\autoref{fig:library-count}). 
%
We conveniently sampled via searching for terms like ``visualization'' and ``gallery'' on web search engines.
We selected galleries using a human-usable computer language (per Fowler~\cite{fowler2010domain}) that consist of an official collection of visual examples rendered from self-contained code.
This excludes SQL-like query languages or dataframe-like APIs for data processing and data transformation. 
Similarly, this filters out natural language tools  (\eg{} NL4DV~\cite{narechania2020nl4dv}) as well as partial-specification DSLs that operate over other DSLs (\eg Draco~\cite{yang2023draco2}).
We excluded benchmark or dataset repositories (\eg{}  vizNet~\cite{hu2019viznet}) which are curated to train or test systems rather than be used by humans.
Our definition of galleries used to identify elements in our survey is provisional: since there was no prior formal definition, we describe the general characteristics of instances regarded as ``galleries'' by their creators and users. 
Similarly, this survey is not exhaustive, but merely develops a baseline for gallery form from which to measure interviewee responses.

\parahead{Interview Studies}
We conducted a semi-structured interview study with (N=\numCreators{}) gallery creators and (N=\numUsers{}{}) with users of those galleries. 
Interviews with creators aimed to answer  \textit{why and how do authors create example galleries?}
Our interviews with users focused on corroborating our findings from creators.

Gallery creators, denoted \cxx{example} and \qtc{quoted}, were recruited via email. They were identified from our gallery survey and snowball sampling.
We interviewed two sets of overlapping authors: \cxx{Jeff} and \cxx{Dominik} both worked on \toolName{Vega} and \toolName{Vega-Lite}, while \cxx{Xiaoji} and \cxx{Ib} both worked on \toolName{Deck.GL}.
All but one participant identified as male. Most creators had at $\geq 5$ years experience designing DSLs or APIs and had designed at $\geq 2$ galleries. 
Users (referred to as \uxx{1} and quoted \qtu{like this}) were recruited via social media.
We sought users with experience using \toolName{Altair}, \toolName{Vega-Lite}, and \toolName{Plot.ly}.
We limited our focus to these tools to narrow our scope and because their conceptual overlap provides points of direct comparison: \toolName{Altair} is a python wrapper for \toolName{Vega-Lite}, while \toolName{Plot.ly} is strongly influenced by \toolName{Vega-Lite}.
We used a screening survey to vary our sample across background and notational experience.
(8/9) have $\geq$2 years experience, (7/9) hold degrees in computing, (4/9) identified as female.
\revised{We piloted the study with two creators and three users. It was marked exempt by the University of Washington IRB.}


We conducted all interviews over Zoom. 
Consent was elicited via email prior to the interview.
Creators received a \$50 gift card in compensation. Users received a \$15 gift card. 
Our interview scripts can be found in the appendix.
Interviews lasted $\sim$60 minutes with creators and $\sim$30 minutes with users.
Demographics were solicited via post-interview survey. 
Interviews were transcribed by the first author \revised{using Otter.ai} and then open-coded to identify the initial themes. The research team iteratively refined these concepts. \revised{Themes were finalized via an affinity diagramming~\cite{luceroAffinity} session. }

\parahead{Positionality}
Several participants have worked with the research team, see appendix for an enumeration of these connections.
In an evaluative study, this would pose ethical quandaries about participants' ability to give unbiased answers. 
We seek to understand and synthesize these expert's experiences.
While we try to circumvent these issues via iterative theming and reflection, our relationships with participants will inevitably bias our results.
Of particular note is our connection with Heer (as colleague, advisor, and collaborator) and his collaborators---for instance he has worked with \cxx{Dominik} (on \toolName{Vega}), \cxx{Fil} (on \toolName{D3}), and \cxx{Matt} (on \toolName{Idyll}).
This centers our results on the approach to tool-making expressed through his work and the (large) surrounding community he has fostered.

\begin{figure}[t]
\centering
\small
\begin{tabular}{|p{0.18\linewidth}|p{0.72\linewidth}|l|l|}
\hline
\textbf{Name}                      & \textbf{Main notation}                                         \\ \hline
Jeff Heer                          & \toolName{Vega}~\cite{satyanarayan2015reactive}                \emph{ grammar of interactive visualizations}                                                           \\ \hline
 Matt Conlen & Idyll~\cite{conlen2018idyll}  \emph{ markup language for online interactive articles} \\ \hline
Nicolas Kruchten                   & \toolName{Plot.ly}~\cite{plotly}                       \emph{ commercial open-source graphing library}                                   \\ \hline
Carlos Scheidegger                             & \toolName{Quarto}~\cite{quarto}                        \emph{ open-source scientific and technical publishing system}                          \\ \hline
Xiaoji Chen                        & \multirow{2}{\linewidth}{\toolName{deck.gl}~\cite{deckgl} \emph{WebGL-powered framework for large-scale data visualization}}
\\ \cline{1-1} Ib Green &
\\\hline 
Josh Sunshine                      & \toolName{Penrose}~\cite{ye2020penrose}                       \emph{ platform for creating diagrams illustrating concepts}                                     \\ \hline
Dominik Moritz                     & \toolName{Vega-Lite}~\cite{satyanarayan2016vega}                     \emph{ high-level grammar of interactive graphics}                                \\ \hline
Philippe \newline Rivière  (Fil)                               & \toolName{D3}~\cite{bostock2011d3}                            \emph{ low-level JavaScript library for interactive data visualizations}                                \\ \hline
Will \newline Crichton                                & \toolName{Nota}~\cite{crichton21Nota}                           \emph{ a language for writing modern documents}                      \\ \hline
Peter Vidos                               & \toolName{ipyvizzu}~\cite{ipyvizzu}                            \emph{ animated charting library}             \\ \hline
\end{tabular}
\vspace{-1em}
\caption{We interviewed \numCreators{} gallery creators or maintainers. We include their de-anonymized names (with consent) to credit them for their work and insights.   
}
\label{fig:creators}
\vspace{-2em}
\end{figure}


\section{What are galleries?}
\label{sec:galleries-what}

\begin{figure}[t]
\centering
    \includegraphics[width=.9\linewidth]{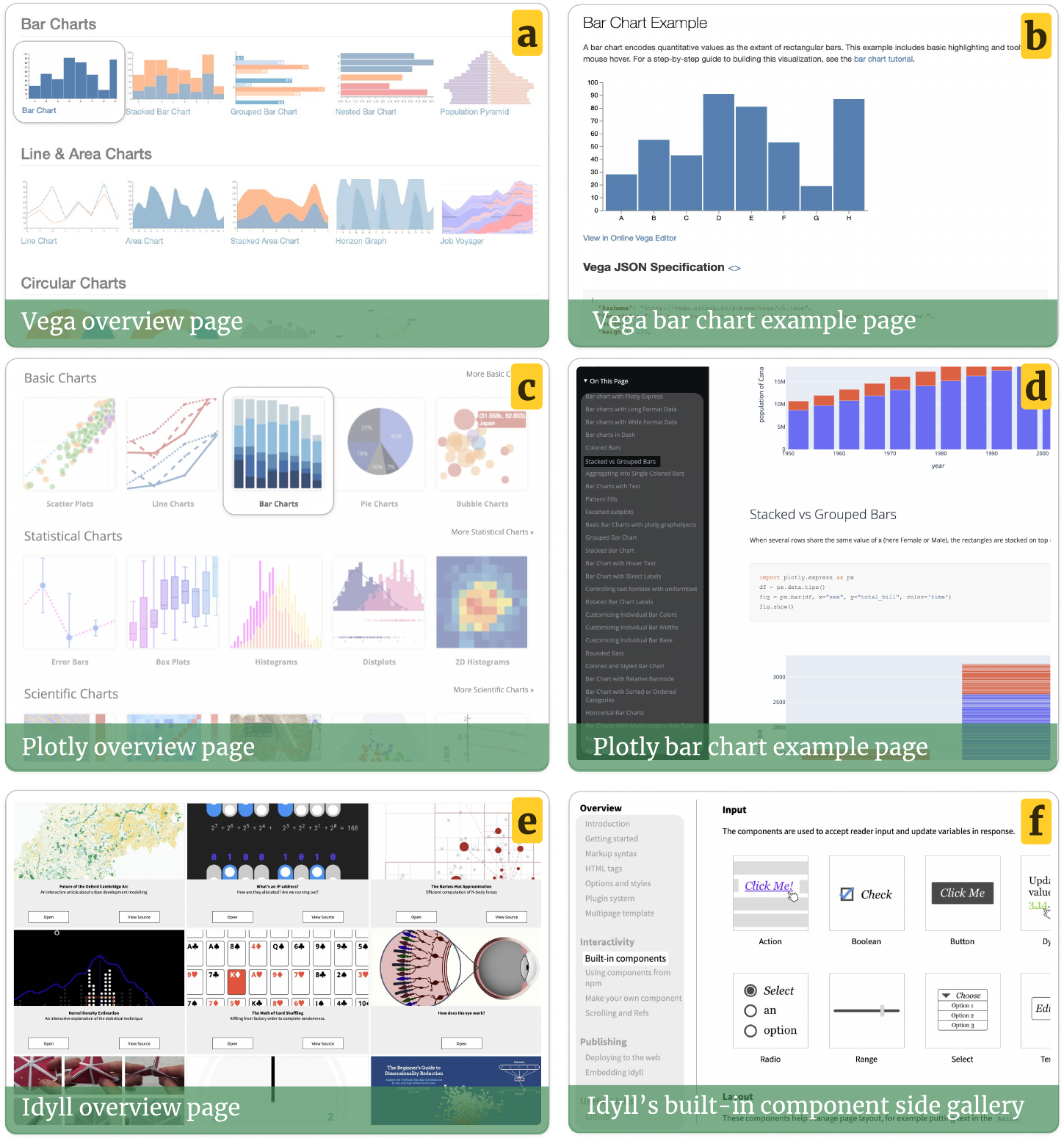}
    \vspace{-1em}
     \caption{\toolName{Vega}, \toolName{Plot.ly} and \toolName{Idyll} Galleries. 
     Gallery organization and content depend on the goals and scope of the project they describe. 
Larger galleries are often categorized by features or task. To wit, \toolName{Vega} groups the examples as bar chart, line charts and so on, whereas \toolName{Plot.ly} groups by data domain like finance or statistics. In contrast \toolName{Idyll}'s gallery is bifurcated into large high-level article examples and small low-level component examples.
     }
    \label{fig:vega-plotly-idyll}
        \vspace{-1.5em}
\end{figure}

As with many things in engineering, there is no one definition of a gallery. 
The type of galleries employed seems to be informed by library nature and construct. 
We identified three forms of example galleries: \textit{showcase} (analogous to a carefully curated museum gallery, in which each piece is a demonstration of a realistic use case), \textit{reference} (analogous to code examples in API references that are minimal and focused), and \textit{index} (analogous to a phone book or table of contents). 
Each library contains at least one \textit{main gallery} that can typically be accessed from  the navigation bar. For example, (a) and (c) of \autoref{fig:vega-plotly-idyll} show small slices of \toolName{Vega} and \toolName{Plot.ly}'s main galleries.  
More complex and customized examples appear either as a ``showcase'' or ``miscellaneous'' category in the main gallery or as an independent gallery focused on that topic (such as \toolName{ipyvizzu}), depending on the number of examples. 
Some libraries incorporate a smaller \textit{reference gallery}
that resembles the \textit{references} type of documentation: it details the lowest level breakdown of library usage.
However, unlike the API references where the textual information appears first followed by an optional code example, the reference galleries are dominated by leading visual examples followed by explanations. 
To wit, \toolName{Idyll}'s~\cite{conlen2018idyll} main gallery (\autoref{fig:vega-plotly-idyll} (e)) accommodates interactive articles, whereas its ``built-in components'' section (\autoref{fig:vega-plotly-idyll} (f)) exhaustively lists the components orthogonal to each other in terms of functionality. 

Similarly, creators did not express a cohesive opinion on what constitutes a gallery.
Some seemed to regard them as the interface to a database of self-contained examples (\toolName{Matplotlib}), whereas others saw them as more bespoke curated collections (\toolName{deck.gl}). 
This heterogeneity can lead to strong opinions, for instance \cxx{Nicolas} argued that \toolName{Plot.ly}'s ``Examples'' webpage (\autoref{fig:vega-plotly-idyll} (c)) is not a gallery, noting that the thumbnails are not selected from their examples, but are rather visually appealing images representing the themes of visualization collections. 
In NotaScope~\cite{kruchten_metrics-based_2023}, \cxx{Nicolas} and co-authors define galleries as a way to evaluate the expressive or generative power of a notation. While it conforms to one of the goals for galleries in our findings, we extend the definition and regard \toolName{Plot.ly}'s ``Examples'' webpage as an index gallery. 
From these views we synthesize the following definition: \\
\textbf{{
    {A gallery is a (potentially) organized \emph{collection} of examples, which consist of code, visual output, and metadata, and an \emph{interface} through which to use those examples.}
    }}

\section{Gallery Motivations}
\label{sec:gallery-values}
Here we identify motivations guiding gallery design such as using the gallery for marketing,  as objects for quick reuse, and as a component of the system infrastructure.

\parahead{Gallery as Marketing or Advertising}
An example gallery is often the face of the library, acting as an ambassador for the library as a whole.
Rich evocative galleries can serve as marketing materials by demonstrating the design space of possibilities for the library, possibly attracting new users or helping prospective users differentiate from similar offerings. 
While the design space is interpolated from the \textit{collection} components, effectively communicating it depends on the \textit{interface} component.
\cxx{Matt} observed that \qtc{people come to the site, and the first thing $$[is that]$$ they'll scroll through the examples to see `if I were to try to make something like this? Is this up to my quality standards?'}
Galleries designed with motivation emphasize demonstrating expressiveness and unique facilities of the language---a signal of development status and evidence of community---as well as sketching what is possible with a notation---so as to elicit adoption or usage.

This interest in marketing affects the selection of gallery elements. \cxx{Jeff} and \cxx{Fil} noted that some highly customized examples are aimed and are not expected to be frequently reused. 
For instance, a community contributed example shows that it is possible to construct a working Pac Man game in \toolName{Vega}, which can serve to highlight the power of the library and promote interest in it. Such examples are added to show the upper bound of what it can achieve as a testament of power. 
Similarly, \cxx{Matt}, \cxx{Josh}, and \cxx{Xiaoji} noted that they consider \textit{aesthetics} to be an important factor. To this end \cxx{Josh} noted \qtc{we have other examples that are available in the source code repository in GitHub. But they’re not as visual appealing or widely useful. So the gallery is intended to be a subset of those that are visually appealing and also representative of the kind of things that we expect our users to want to do}. 
Besides the official gallery, to help answer users' questions \cxx{Fil} also maintains a personal collection of Observable notebooks, and creates one-time examples that are archived online afterward. 
These suggest a multi-tier example database with a front-facing curated gallery and a repository for additional examples. 


Commercialized libraries such as \toolName{Plot.ly} put in enormous effort to compete for market share, and often leverage their gallery not only for marketing its features, but also as a means of Search Engine Optimization (SEO).
This is done with the explicit goal to be at the top of search results for keywords like (\eg ``Python scatter plot''). 
\cxx{Nicolas} explained that the strategy involved ensuring that \qtc{every feature gets an example...It's not something that's meant to be browsable as a unit; it's meant to be linked directly to any sort of scroll up and down and optimize for using CMD+F within a page or Google or our search feature}.
For example, \uxx{2} started exploring \toolName{Plot.ly} after finding a StackOverflow answer implementing the design they need (\eg{} Kernel Density plot with zooming). 



An important usage in this advertising situation is supporting comparisons with other prominent tools---potentially helping users answer the question \emph{why should I use this tool instead of another?}
For instance, \cxx{Jeff} observed that \qtc{galleries from $\toolName{D3}$ will typically involve lots of visualizations that you can't make with $\toolName{ggplot2}$, or $\toolName{Observable Plot}$ or with $\toolName{Vega-Lite}$, because they target different levels of expressiveness. And so as you start to do some of that sort of comparative research across tools, you will also get a sense}.
Multiple galleries together help users select a suitable library by comparison~\cite{tanzil2024people, kruchten_metrics-based_2023}. 
Users recalled needing to use a tool's gallery to \qtu{figure out what it can and cannot do} (\uxx{2}) and \qtu{learn the visualization vocabulary} (\uxx{1,3,6}) when they started to learn the libraries. 
User participants suggested that they tended to select tools based on the interactions the library supports (\uxx{1,2,4,6,7}), ease-of-use (\uxx{1,5-7}), and its popularity (\uxx{2-7}).  
Gallery's can signal these traits, such as by showing GitHub stars, impressive quick starts, or just by showing the gamut of possible interactions.


Users consider adopting a library based on certain qualities that facilitate their specific goals. 
For instance, in comparing \toolName{Plot.ly}, \altair{} and \vl{}, our user participants expressed a variety of usage. 
These include analyzing datasets (\uxx{1,2,4}), presenting findings (\uxx{1,2}), academic publishing (\uxx{1,3,5,6}), testing research prototypes (\uxx{1,3}), and teaching data science classes (\uxx{1,7}). 
This is then sometimes reflected in gallery organization, for instance \toolName{matplotlib} explicitly includes sections focused on various domain tasks relevant to those roles.  
Some differentiating features can be simple. For instance, \uxx{3} integrated \toolName{Vega-Lite} to generate SVGs used in their research prototype. While other tools support that output format, \toolName{Vega-Lite} did so more explicitly.



\parahead{Gallery as Reusable Template Corpus} 
\label{sec: reuse}
Another important motivation for gallery construction is to help users overcome the blank page problem, such as by providing inspiration and facilitating reuse.
Gallery \textit{collection} determines the existence of reusable contents, whereas its \textit{interface} affects the effort to locate them. 
Galleries developed with this motivation seek to provide ideas for common tasks (\eg{} making quotidian chart types like bar charts) as well as design alternatives by displaying the possibilities. 
\cxx{Jeff} observed that a gallery can \qtc{give them an idea for visualizations they might not have considered otherwise. And so it's part of showing what's possible, and the way it might inspire someone.}

Some creators believe that examples are designed to be repurposed, functioning almost like a slow auto-complete.
\cxx{Dominik} added that this use of examples is similar to templates~\cite{bako_user-driven_2023, mcnutt_integrated_2021}, noting that  \qtc{Galleries are just fairly lightweight way to get lots and lots of different programs that show different aspects and different patterns. And hopefully provide good starting points for people to copy and modify. So they don't have to start from scratch}.
This follows a pattern known as opportunistic reuse~\cite{brandt2009two},  in which users copy and paste code, and then modify it to meet their needs.  
Users expressed exactly this behavior, explicitly noting that they seek examples for reuse when they have a target chart type in mind. 
For instance, \uxx{5} browses the gallery examples looking for additional accessories when they have a base design. \uxx{3} is familiar with the gallery and has a mental map of each chart's location, so they can easily jump to the example needed and use it as starter code. \uxx{4} prefers to view the examples as an aid to iterate over the design they brainstormed: \qtu{so it's more of a learning process...there were visualizations that you did not imagine, but now that you know them makes more sense than the things before. }
When using tools like \toolName{D3} or \toolName{ggplot2}, they often search for more elaborate designs from sources like {``Information is Beautiful''}~\cite{information_is_beautiful} (\uxx{1}) and search engines (\uxx{5}). 

Galleries do not serve every (re)use case---with folk documentation sometimes being more accessible. For instance, when using \toolName{D3}, \uxx{5} prefers searching online to the gallery because searching is likely to direct them to Stack Overflow where a lot of questions are directly answered. 
Similarly, \uxx{1,2,6,7} added that they make common charts types (\eg{} line charts) from scratch because it only requires a few lines of code from memory. They refer to the gallery when in need of examples that are \qtu{not the simplest, like anything stacked} (\uxx{6}) and when \qtu{it's just easier to start from something exists than from scratch} (\uxx{3}). 
Similarly, not every gallery is built to service this style of usage. \cxx{Nicolas} was skeptical about the benefits of opportunistic reuse in reality for smaller galleries (\eg \toolName{Seaborn}): \qtc{the likelihood that one of these things happens to look like what you're trying to make is pretty low. I mean, the design space for charts is huge, right? I just think that the browsing approach to like, providing this information to users doesn't scale.} 
Opportunistic reuse is less likely to be applied to showcases and highly-customized examples since they also require the users to interpret the code, decompose and extract relevant pieces. 







\parahead{Gallery as Test Infrastructure}
Finally, some creators were motivated by using gallery examples as part of their test infrastructure, namely as unit tests. 
Unit testing is an established practice motivated by developers' conviction and management requirements to ensure software quality~\cite{daka2014survey}. However, programmers often look for usage examples from unit tests, when there are no better-documented options, even though it can be time-consuming~\cite{nasehi2010unit}. 
In contrast, galleries contain self-exploratory usage examples mainly documented for the users, but they benefit the creators as end-to-end test cases. 
Compared to low-level unit tests, they give an opportunity to check how well things work when fused together (\cxx{Jeff}). 
They help developers quickly validate the ideas, serve as a debugging tool for local development via visual inspection (\cxx{Xiaoji}), and aid in the discovery of  performance issues (\cxx{Matt, Ib}). 
While the testing is mainly carried out through the \textit{collection}, the \textit{interface} provides a means of inspection. 
\revised{Similarly, while this approach does not directly benefit the end users, it can affect the perception of the library, such as via coverage badges and commit activity (as signal for stability).}

However, using galleries for testing requires setting up testing infrastructure in addition to the general gallery maintenance. 
This can be beneficial: \toolName{Penrose} guarantees every example in the repository keeps running using continuous integration (\cxx{Josh}). 
The granularity of these tests can vary. For instance \toolName{Vega-Lite} and \toolName{Penrose} programmatically check the rendered SVGs for each gallery example against snapshots to assure consistency across code changes, while \toolName{Quarto} merely checks that the documentation site successfully loads and produces documents using browser automation. To refine the testing pipeline, \cxx{Carlos} planned on inspecting the example contents \qtc{to not only generate output, but also assert things about the output.} 
However, the utility of automation has limits, with manual inspections still being unavoidable in some cases. For instance, the effect of code changes can be subtle and not visually noticeable (\eg slight changes from \toolName{Penrose}'s layout optimization algorithm), or they can be significant and still correct by definition. As a result, they require human intervention as part of the testing process. 
\cxx{Jeff} mentioned checking if all the charts looked correct before a big release. 
Narrowing the gap of evaluation between programs and their intended interfaces/outputs seems to be a crucial but also incredibly challenging goal for the creators to achieve, and the gap seems to widen as libraries become more expressive.

\begin{figure}[t]
    \centering
    \includegraphics[width=\linewidth]{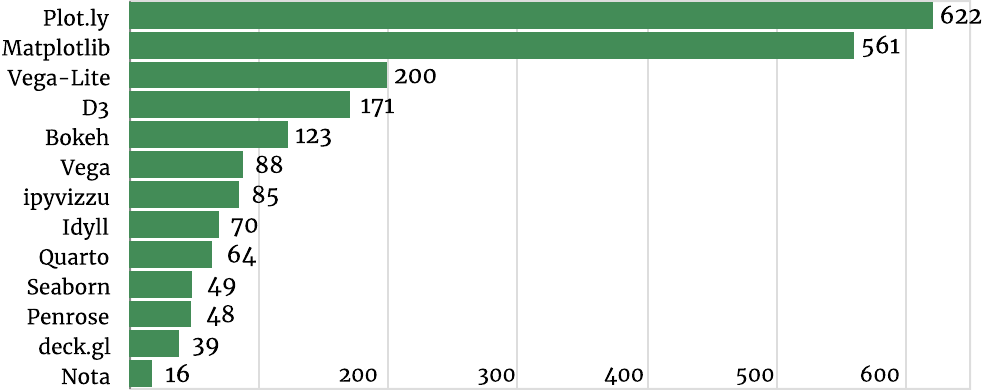}
    \vspace{-1em}
    \caption{
    Example counts in galleries we explored to situate our analysis.
    }
    \label{fig:library-count}
            \vspace{-1em}
\end{figure}

\section{Cross-cutting Themes}
\label{sec:gallery-themes}

We now explore several cross-cutting themes from across the gallery and their relationship with different motivations.

\subsection{Education and Communication}

Galleries are often the main point of contact between users and creators, meaning they are laden with communicative tasks, such as showing how to use a library. 
\cxx{Nicolas} emphasized that \qtc{if it's not in the docs, you may as well not have built it. No one is ever going to use it if it's not in the docs.} 
Here we discuss issues related to communication of functionality.




\paraheadd{Are galleries a teaching tool?}
Some creators saw galleries as a teaching aid that helps engage and educate people in exploring the library. \cxx{Jeff} noted that while they lack the detail of a tutorial, galleries allow users to see how things are constructed and how goals are achieved with the library.
In contrast, \cxx{Josh} expressed a marketing-centric view,  arguing that the gallery is not the best place to teach a library. 
He argued that \qtc{the main purpose of the galleries is an advertisement, not an explanation or something that people are intended to reuse.}
\cxx{Matt} echoed this perspective, noting that while people learn from examples, but \qtc{if you want to teach something, you want a simple example that highlights the thing you're trying to teach, which are not usually good fits for the gallery because of the visual complexity and their applicability to a lot of people}. 




The user participants all started their journey with the charting libraries by playing with the gallery examples. 
As beginners, they were not \qtu{trying to become an expert, but just wanted to use it} (\uxx{5}). They browse the overview page and select basic examples to replicate. Then they slightly tweak the examples by changing the parameters, and that was when they refer to the references for detail. 
In addition, \uxx{3} highlighted that he learned about visualization research and vocabulary by studying \toolName{Vega-Lite}'s gallery. 
We infer that the effectiveness of gallery as a learning tool depends on gallery structure, library construct, and the nature of the domain. Galleries of charting libraries are broadly used as learning tools because they are often categorized based on well-studied taxonomies, and provide basic and straightforward examples. In contrast, \toolName{Idyll} and \toolName{Penrose} have no such categorization to lean on.

Examples are sometimes used as a means to educate about a particular topic.
\cxx{Jeff} sought to have \toolName{Vega} gallery examples to convey useful data facts in addition to features of the system.
Correll~\cite{correll2019ethical}, referring to example data more generally, calls for selecting datasets that are of social import or impact as means for advocating for positive change.
We suggest that this usage of examples is under-explored and may benefit library marketing (via novel examples) as well as social good. 

Crichton~\cite{crichton2020documentation} (\ie{} \cxx{Will}) argued that documentation generation can be seen as a form of information visualization.
We connect this observation with Lee-Robbins'~\cite{lee2022affective} characterization of visualization as being a means of fulfilling learning objectives. 
We argue that galleries are then, in essence, a visualization of what is possible in a gallery in a means that is meant to support learning of those features---whether or not that learning is intentional. 
\revised{Although the creators and users agree on the tutorials being a more conventional teaching material for the beginners, gallery examples provide shortcuts for more focused further learning. }
We suggest that developing designs to help users bridge from gallery use to tutorial usage is an important documentation challenge, and may be an area where new tools may be helpful.



\parahead{Community Examples}
Eliciting external examples is a commonly used means to expand galleries. 
For instance, after making an initial collection of showcase examples for \toolName{Idyll}, \cxx{Matt} gathered more instances in the form of student projects from a course he taught at University of Washington. 
\cxx{Jeff}, \cxx{Dominik}, \cxx{Carlos}, and \cxx{Josh} echoed the value of selecting existing external examples as gallery entries. 
To this end, some libraries (\eg{} \toolName{p5.js}~\cite{p5js}) specifically facilitate new users' ability to contribute examples (and other forms of documentation), and try to make the process of adding to that documentation a welcoming and easy process.

In some cases external community resources form to support folk curation of these examples, such as on StackOverflow or in a dedicated website like \toolName{bl.ocks.org}.
The latter of these is a (defunct) means for the \toolName{d3} user community to collect and browse examples. 
Making a formal connection with this type of folk resource is a potential area of tool support.
\cxx{Dominik} envisioned a two-tier system where a maintainer can \qtc{check a box} for a community example to promote it into the official gallery. Forked versions could be tracked to show variations---perhaps inspiring a remix culture \ala{} OpenProcessing~\cite{subbaraman2023forking}. 

Community examples are not a silver bullet. Despite the example creation labor being outsourced, the curation and management overhead for quality control still apply to monitoring user-contributed examples. 
Similarly, \cxx{Jeff} noted that they can muddy the utility of a tailored collection of examples.
We note that this tends to cause these examples to be more like showcase examples (which is how \toolName{Quarto} uses them), which can limit their re-usability.  
Imposing strong restrictions may support higher quality examples, but can also severely reduce the number of community submissions to the gallery.




\parahead{Communication \& Expectations}
Barriers exist both in delivering information from the creators via documentation and receiving questions and feedback from the users. 
While a number of channels are available (ranging from the galleries themselves, to secondary channels like GitHub or Discord), designing an appropriate mixture of communication forms is dependent on the community around those tools. 



The inherent knowledge asymmetry between users and creators makes it difficult to form and select examples.
For instance, \cxx{Carlos} highlighted that \toolName{Quarto} has a large and diverse set of users with varied technical background, noting \qtc{our examples use \textsc{pandas} because we expect it to be a familiar tool for Python users, but it's a very real problem that we have to guess what's the knowledge base, and have different ways to explain things in code}. 
User participants struggled with integrating multiple tools (\uxx{6}) and spent considerable time adapting data to match visualization specifications (\uxx{1-7}).

Some creators interact with users on social media platforms to answer questions, collect feedback, and observe how users review the tool. 
Most users would not interact with the creators directly, however, they typically find answers to their questions through StackOverflow or GitHub issues (\uxx{1,2,4,6}), and reviews containing research done by others (\uxx{2}). 
When users express confusion about gallery examples in submitted GitHub issues,
\cxx{Xiaoji} adds more explanation to the \toolName{Deck.gl} examples. 
In these interactions users contribute feedback that can help shape the tool collaboratively in manners that increase its usability (supporting both marketing and reuse goals).
They can also act as a slow form of testing by communicating the pain points they encounter with the tool.
To this end, one surprising channel of communication from user to creators is through user-contributed examples.
For instance \cxx{Carlos} uses the community contributed \toolName{Quarto} showcase as a way to monitor how people use the tool:
\qtc{a number of the things that people have built are not things we would ever have considered. There's some very advanced things that people do with styles that are really nice}.
Correctly positioning the role of contributed examples can then serve a dual purpose of providing documentation and soliciting feedback. 






Communicating correct usage patterns is essential, both to ensure that users understand what the notation can do (as in the comparative research use case of marketing), as well as show how to do those things. Ineffective articulation of these patterns may lead to user confusion. 
For example, mismatches in example-suggested programming patterns and the user's target implementation can hamper code reuse, as observed by \uxx{1}:
\qtu{sometimes they wrap the visual functions in a class, and it's hard to extend to some edge cases.} 
\cxx{Ib} observed different strength and hurdles for community in different groups for developers familiar with functional paradigm, which \toolName{Deck.GL} uses, or imperative programmers. It is intended to be used by web developers who are familiar with React and its functional paradigm, and can be confusing to other groups. In contrast, React users can have their own problem: \qtc{it's very hard to combine React with an imperative API, which $\toolName{Deck.GL}$ $$[often needs and]$$ works extremely well with} (\cxx{Ib}). 
The usability of code examples and documentation is affected by how the notation is designed, just as other parts of technical systems~\cite{jakubovic2023technical}. 
Creators make assumptions about usage patterns and work backwards to inform their design decisions. While these decisions might benefit the \emph{assumed} user group, they could make it less flexible for less expected or expert users.

Similarly, concepts from within the domain are difficult to capture unambiguously.
For instance, in selecting appropriate names for examples, \cxx{Nicolas} observed that
\qtc{It wasn't super clear, what would you call things. Like do you call things facet and sort of trellises or small multiples or panel charts or lattice? I don't know. So we tended to kind of go with everything, like, make sure we can stick all the keywords in the page.}  
Communicating usage patterns is extremely difficult for libraries supporting interactions and animations because they add extra dimensions to the design space, which also requires more domain knowledge. 
For instance, \uxx{4} complained about the difficulty of identifying valid combinations of static design and animation operations.
Users care about the end-results rather than the steps that lead to them, however, the space of possible usage patterns can be large, which can make searching for specific solutions challenging. 
\emph{Instead of merely enumerating the options (via the gallery), generative recommenders might offer result tuned to user questions.}

\subsection{Maintenance}
Gallery creation and maintenance is a huge burden.
Here we review several challenges latent to this process.

\parahead{Resource constraints} Growing a gallery can be extremely demanding without immediately demonstrable benefits.
Moreover, it is unclear how to most effectively distribute limited developer time. Should galleries be weighted above over tutorials? Is new documentation or a new feature more important?

Weighing changes to the maintenance workload against the library goals is one means to determine resource allocation. 
For example, \toolName{Plot.ly} explicitly focuses on Search Engine Optimization (SEO), likely in part because they have such strong marketing centered goals, and so the costs of maintenance are offset by the potential value of increased SEO. 
In contrast,  \cxx{Ib} believed in spending more time building great features to attract users rather than add more examples and improve the documentation. 
Libraries like \toolName{Seaborn} are led by a few part-time developers with \qtc{probably too little time to consider the best things to do with galleries} (\cxx{Nicolas}). This contrast can be seen in \autoref{fig:library-count}: \toolName{Plot.ly} has nearly 13$\times$ more examples than \toolName{Seaborn}. 
Is this a better allocation of resources? It depends on the intent of the library.


Creating a gallery is usually easier at the beginning as presenting straightforward use cases that justify the need for new features in a library. 
Although, after developing new features, \cxx{Ib} described how they implement the corresponding examples, noting that they \qtc{have to do all the extra work of actually create a beautiful example or finding a dataset that's public with good license. So sometimes it's just like an extra tax on the work to do it}.
Further, {as a gallery grows, it requires a series of careful considerations to select and organize the examples}. 
Additional interface design (such as search bars, categories, or other affordances) for the gallery becomes essential for efficient example discovery, but it can introduce more cognitive overhead.  
\cxx{Josh} explained that \qtc{if you have a category or directory or a search, then you have to maintain that. It's another feature to deal with. And it adds complexity to the user experience a little bit, because it's another thing they have to go through in order to get started. So there's no reason to do it till you have to}.
This may suggest that tools to support the automated construction and organization of galleries would be useful, as it would relieve some of this burden. 
However, the prominence of auto-doc tools like Sphinx~\cite{sphinx} would suggest that this problem is not as simple as building a new documentation system.

\parahead{Maintaining and Coordinating Artifacts}
There are multiple artifacts around galleries that require coordination. These include the gallery examples, software features, testing suite, and the software ecosystem more broadly. Actions on any of them would likely cause consequences on another component.
For example, \cxx{Dominik} described that when a new example is added to \toolName{Vega-Lite}, it is required to be included in the Vega Editor (\ie online playground). 
Further, if there are changes to an example's data, the \texttt{vega-datasets} repository needs to be updated, which in turn involves ensuring that other examples do not break. 
Notebooks such as \toolName{Observable} provide a web-based platform for hosting customized examples written in all sorts of libraries a user can import. \cxx{Fil} mentioned how using \toolName{Observable} reduced the workload for managing the gallery interfaces of \toolName{D3} and \toolName{Observable Plot} since the examples are just links to notebooks. 
However, making a system seamless for someone inevitably makes it frictional for someone else~\cite{hengesbach2022undoing}. For instance, moving documentation into notebooks (or Discords in some cases) can make that documentation less discoverable (\eg{} for search engines) and potentially more intimidating to some users.


All creators expressed that it is challenging to maintain galleries when the core software changes. 
\cxx{Matt} explained that \qtc{maintaining documentation for an open source project is a real drag for the most part}, going on to note that \qtc{every time you do that \/[fixing bugs, updating APIs, \etc \/] you have to go and update the documentation too. So it's like this added layer of work, and it becomes hard to keep things in sync.}
Manually inspecting each example is a common task before releasing. 
\cxx{Jeff} mentioned going through examples manually as a sanity check before big releases because not all examples have a one-to-one mapping to testing and validation. 
\cxx{Josh} reflected on the experience of having to re-implement every example after updating \toolName{Penrose}'s features, noting that even if everything works well, \qtc{\/[adding more examples\/] adds time to the testing \/[because\/] we run a test suite every time we commit something to the repository.} 
\uxx{5} disliked how software version updates can affect the community examples, potentially breaking examples and confusing readers (particularly of unofficial documentation, like on StackOverflow). 
\cxx{Nicolas} stated that software with a larger and more stable audience (such as the marketing-centric \toolName{Plot.ly}) enforces a stronger commitment for backward compatibility and instead \qtc{treats everything as the documentation problem, because \/[the library\/] is pretty hard to evolve}.
Newer libraries are likely to go through major changes, requiring additional labor to fix the examples. In contrast, mature libraries tend to keep everything as is, which ensures stability but can impede new features.

As a library phases into the maintenance period from active development, the frequency of adding new examples also slows down. The creators occasionally get inspirations and implement new use cases. Yet more often, enthusiastic users contributed a burst of examples in short period of time as the user base grow (\cxx{Xiaoji}). Another way to create new examples is demand-driven in that multiple users asking about a similar question indicates the need for a new example \cxx{Nicolas, Jeff, Fil}. 
Library creators and maintainers answer questions by showing examples. And some of the examples become part of the gallery, while others are never revisited by the authors (\cxx{Fil}).
Maintenance is typically invisible labor~\cite{akbaba2023troubling} that rarely garners the attention that new feature work does, but may offer a marketing-style signal that the library is actively maintained.
\subsection{Gallery Contents}

Finally, galleries are agglomerative objects. They consist of many constituent objects, the organization and selection of which is guided by balancing different views and constraints.



\paraheadd{What's an example?} The main contents in an example are the caption, the visual, the explanation, and the code. 
The galleries all contain these elements but with different levels of detail. 
For example, a detailed chart example can incorporate the background of the data source and an explanation of why the visual encodings were selected to be more educational.


Creators often took inspiration from other galleries in the selection and design of their examples. 
This expresses a marketing-tact: it allows them to show that they have the parity of what others can do and more.
For instance, being competitive for market share, \toolName{Plot.ly} reproduces examples from \toolName{ggplot2} and \toolName{Matplotlib}, \qtc{literally saying `do you like this $\toolName{Matplotlib}$ plot and here's exactly how you make it'} (\cxx{Nicolas}).
This is akin to how a new library is also a reaction to the flaws (\ie negative space) of its alternatives, which \cxx{Ib} highlighted as being especially important.
\cxx{Dominik} noted that these considerations started from the beginning of \toolName{Vega-Lite}:
\qtc{then we have some examples of showcasing different things people often want to do. And that's how we actually designed language: we made examples and invert the language $$[from the examples]$$ so that it works with the examples that we designed}. 

Creators become more selective as more examples are added (\cxx{Matt}), and  they ensure that the difference of a new example from existing ones has a clear target in functionality or learning objectives by carefully designed ordering (\cxx{Jeff}). In general, creators avoid repetitive examples that overwhelm users by the quantity. 
\cxx{Jeff}, \cxx{Dominik}, and \cxx{Josh} noted that try to keep the examples intriguing while filtering out niche examples that distract users with irrelevant details. For example, \cxx{Dominik} noted that new \toolName{Vega-Lite} examples reuse datasets from previous examples when possible to share data domain knowledge across examples.
\cxx{Ib} observed a related tension between performance and understandability, noting that: \qtc{to make sure they load fast we apply some compression on those datasets, and the example code is no longer easy to understand by the end user.}
Selecting examples requires balancing their need to interesting, understandable, re-usable, and to not unduly  burden the infrastructure.

\parahead{Gallery Organization} The order in which gallery examples are presented and the means by which they are organized strongly affects the readability.
Most galleries navigate this challenge by sorting and clustering by complexity or similarity.

The galleries tend to start with simpler examples and progress to more specialized or complex cases (\eg, layers and interactions in \toolName{Vega-Lite}), both throughout the entire gallery and within each category.
\toolName{D3}, as a special case, arranges the more complicated categories such as animation and interaction at the beginning and the basic charts at the end. \cxx{Fil} explained that they intentionally arrange the gallery in this way: \qtc{it serves the users better to use $\toolName{Observable Plot}$ (a high-level charting library by the $\toolName{D3}$ creators) because they will have all the defaults that help them create basic charts.}
Conceptually and visually similar examples are placed closer together to avoid losing track or confusion. 

Several other strategies were prominent as well.
For instance, \cxx{Carlos} explained that they would shape \toolName{Quarto}'s components around how they present the usage in the documentation. If a feature does not fit well with the narratives they will redesign it to match the concepts behind. 
In contrast, \toolName{Plot.ly}'s emphasis on marketing affects how their gallery is organized: an extensive set of design variants with descriptive headings are stocked in the same webpage, so that the webpage can be better indexed by search engines (\ie{} for SEO).

Users have a finely tuned process for sorting through examples and deciding relevant instances. They look at different cues in the environment, such as the example titles and the code structures. Even when they look at unfamiliar galleries, they are capable of interpolating from their experience with other libraries or programming in general. 
\vl{} and \altair{} users (\uxx{3-6}) browse the galleries when searching for relevant examples, while two of the \plotly{} users (\uxx{2,7}) start from Google searches as expected by the gallery designers. \uxx{3} found the categorization useful for quickly locating the charts, compared to \uxx{6} who is not as conscious but became aware of the organization:
\qtu{I have seen that page a fair amount of time by now. So I know how it's organized. I know that the slightly crazier things and the maps are at the bottom. }
\uxx{1} spoke about how they wish there were an additional overview of all variant examples with a \qtu{more scientific} designed arrangement---highlighting the potential value of design interventions for helping users understand gallery contents.



\parahead{Gallery Peripherals}
 Finally, augmenting this gallery organization and content are a family of other tools, that provide context, support inter-connections with the rest of the documentation, and bridge gaps left by the default presentation.

Besides common elements like text and image, a few galleries employ cross-links to references, tutorials and other relevant examples, and tags of keywords.
A playground or online editor is a common documentation component for external DSL-style libraries~\cite{fowler2010domain}  (\eg{} \toolName{Vega/Vega-Lite}, \toolName{Penrose}, \toolName{Idyll}, \toolName{Nota}, \etc).
These tools allow users to play with the examples and make modifications in-situ.
For instance, the \toolName{Vega} and \toolName{Vega-Lite} editor~\cite{vega-editor}  includes specialized affordances for those libraries, such as debugging~\cite{hoffswell2016visual} and performance~\cite{yang2023vegaprof} tools. 
Similarly, it supports link sharing so that specifications, examples, and demos can be distributed.
This design supports both conceptual reuse---as it make it easy to make adjustments to existing examples---as well as marketing---as it can demonstrate the facilities of the library more clearly than static images of gifs.
Secondary tools that are woven into the documentation can help support its usage and is an opportunity for tool research.

Galleries are one of the many tools utilized in user's workflows. 
For instance, \uxx{6} processes data with Pandas, generates the base chart in \toolName{Altair}, and uses Adobe Illustrator to modify the aesthetics and add annotations. 
\uxx{3} also switches between \toolName{Vega-Lite} and other tools so that he repurposes \toolName{Vega-lite} generated SVGs to the data storytelling usage scenario by adding annotations and animation. 
Most galleries only provide succinct examples which required less reading and code decomposition, leaving a gap for integrating the library with other environments.  
Many users employed AI-assistance as a component of their daily workflows.
For instance, \uxx{1}, \uxx{2},\uxx{4}, and \uxx{5} mentioned using ChatGPT, Phind, and Copilot either in combination with the galleries or as the replacement for specific tasks. 
\uxx{4} explained that they first inspect the code recommended by Copilot based on previous code (\eg data manipulation code) or comments. 
\uxx{2} turns to Phind as a replacement for Google search, while \uxx{1} and \uxx{5} use ChatGPT for implementation and debugging tasks.
Framing the Google search queries gets \qtu{progressively and exponentially more difficult} when they are in later stages inof a project to \qtu{get a bit more creative to format the search} (\uxx{2}). 
With LLMs, \uxx{2} can \qtu{explain more about what $$[they are]$$ trying to do}, even when they could not figure out an appropriate query \qtu{as a non-native speaker $$[in English]$$}. 
Users encountered well known bugs with these tools. 
\uxx{5} complained about the model mixing up syntax versions or hallucinating invalid code, especially for newer or less popular libraries, forcing them to \qtu{spend more time on fixing the errors and then realize that the code doesn't work at all.} For the same reason, \uxx{1} choose \toolName{Matplotlib} as the target library when using ChatGPT. 

Clearly, using AIs to support documentation is becoming increasingly common. 
To wit, many documentation sites now including chatbots as a sibling feature to traditional documentation.
However, we note that this can in turn burden maintainers with disentangling that confusion. 
For instance, \cxx{Fil} observed that \qtc{people try GPT and then they come to us with code that doesn't work. But they don't understand what it's trying to do. And we have no idea how they come to this code and what they are trying to achieve}. 
Moreover, the long term benefit of these generative structures rather than browsable structure is unclear, although we suggest in the future they might be hybridized in useful ways.







\newcolumntype{R}[2]{%
    >{\adjustbox{angle=#1,lap=\width-(#2)}\bgroup}%
    l%
    <{\egroup}%
}
\newcommand*\rot{\multicolumn{1}{R{30}{1em}}}
\newcommand{\XX}{X}

\begin{figure}
\small
\begin{tabular}{ccc|p{0.7\linewidth}}
\rot{Marketing} & \rot{Reuse} & \rot{Testing} &\textbf{Takeaway}\\
    \hline\hline
    \rowcolor{yellow!10}  \XX & \XX && Galleries can bear learning objectives,
    but not all are conceived as the main tool for teaching; setting expectations is critical \\ 
        &\XX & \XX & Community-contributed examples can capture new use cases, but they need to be maintained \\
     \rowcolor{yellow!10} \XX & \XX& \XX& There is no silver bullet for user-creator communication, any seamless solution will inevitably be friction-ful for someone\\ 
     \hline\hline
     \XX &\XX & \XX& 
          Effective allocation of developer time depends on project goals: sometimes improving SEO can be more important than new features\\
     \rowcolor{OrangeRed!10} &&\XX & More examples can capture more use cases, but can add to upkeep\\
     \hline\hline   
    \XX & \XX && Different priorities yield different shapes and sizes of examples, impeding  some expected uses\\
    \rowcolor{RoyalPurple!10} \XX &\XX  && Users often have specific expectations about how to navigate the gallery, which must be considered to ensure legibility\\
    &\XX &\XX& Generative AI can help fill documentation gaps, but can burden communication channels \\
    \hline\hline

\end{tabular}
\vspace{-1em}
\caption{Takeaways by theme: {\hlc[yellow!10]{communication}}, {\hlc[OrangeRed!10]{maintenance}}, and {\hlc[RoyalPurple!10]{content}}.
}
\label{fig:summary}
\vspace{-1.5em}
\end{figure}

\section{Discussion}
\label{sec:discussion}

In this work we analyzed the user and creator perspectives on visualization galleries. 
We find that galleries serve as marketing materials, reusable templates, and testing infrastructure.
Navigating these different usages forms inherently leads to trade-offs and compromises. 
For instance, the organization, structure, and size are shaped by the goals of the creators and the community that surrounding the library. 
We summarize these and other findings in \autoref{fig:summary}.





\paraheadd{Whose gallery is it anyway?}
Creator and user needs from galleries evolve over time, but in different and sometimes conflicting ways. 
For the creators, galleries start as materialization of design space and means to test implementations. 
Marketing and communication needs subsequently arise to attract and keep users, especially when the software transitions from active feature development into maintenance phases. 
In contrast, galleries give users a first impression and serve as the entry point to learning the tool. 
As they get more advanced, galleries become cheat sheets for quick inspiration and reuse. 
Expert users revisit them less frequently because they know the examples by memory or need customized cases from other sources. 
Creators anticipate diminished utility for each new example added to larger collections as they foresee heavier burden for maintenance, the need to redesign, and users' cognitive load.  Unaware of this backstage work, users desire ever larger galleries, being overly optimistic of effectively locating examples in need. 
Reconciling the difference in needs and resources does not have a clear solution, but seems most effectively aided by iterative communication between both groups. 
Beyond this, scaling searching and browsing of examples keeps being an essential interface design challenge.



\parahead{Evaluating Galleries}
Many interviewees stressed the difficulty in assessing gallery quality, especially when isolated from the rest of the documentation.  
For instance, \cxx{Josh} observed that \qtc{It's hard to say how much the current instance of the gallery makes a difference versus just a small number of examples that are highlighting things. It's very hard to tease out all the different parts of the system.}
\cxx{Nicolas} and \cxx{Fil} noted that they keep track of statistics such as webpage traffic in order to analyze what examples are well-received. 
Creators often monitor services like StackOverflow (\cxx{Nicolas}, \cxx{Carlos}, and \cxx{Fil}), Discord (\cxx{Matt}, \cxx{Josh}, \cxx{Dominik}, and \cxx{Fil}), and GitHub (all). 
Creators empirically construct and evaluate the galleries without standard validation mechanisms, which may limit a gallery's potential.
Finding means and metrics to automatically evaluate galleries is mostly unexplored. Kruchten~\etal{}~\cite{kruchten_metrics-based_2023} explore metrics for a particular form of gallery, but their connection to utility is unknown.
We highlight this as a useful area for research.
For example, measuring the cognitive distances between examples might help with better gallery organization and navigation.
Similarly, developing a readability score (\ala{} TuTVis~\cite{sabab2020automated}) for galleries might highlight places for improvement.
A related coverage metric might analyze not just the fraction of the library's API represented, but how much of the essential ideas for its operation are covered.
Realization of such metric might guide both better practice by creators but also automated gallery organization and management.
\parahead{Docs as database}
As we observed galleries can be a variety of different things. One persistent view that we observed across creators, users, and galleries themselves is that they are sometimes viewed as databases. 
For instance, prior work explored gallery-like exploration of visualization collections drawn from databases~\cite{oppermann_vizsnippets_2022, setlur_olio_2023, xu_ideaterelate_2021}.
Extending this metaphor is enticing: it suggests that we can conceptualize ensembles of examples in different collections---as in showcases, indexes, and so on---as views over an underlying data structure. 
Tool support under this metaphor might empower galleries across our themes without negatively affecting the others.
This is a potentially useful venue for AI-assistance. For instance, an example might be defined not by its code, but instead by its task, allowing a generative system to only materialize at the last minute, allowing it to be more contextual and specific to a user's interests. 
However, this may clash with the curation of bespoke experiences that are sometimes intended for galleries---a collection of representative examples of sustainable size. 
Future research could explore malleable galleries that unify these perspectives by ranging between select all-style queries to more bespoke personalized presentations that adapt to user needs and experience, and, ideally, strike a balance between facilitating maintenance and learning.

\parahead{Limitations} Like any work, ours has its share of limitations.
We interviewed a relatively limited set of creators and users. 
A survey would have had further reach, but we instead elected for a smaller number of higher quality responses. Such a survey is useful future work.
Our user interviews are influenced by our selection process and may not capture the usage patterns of those who do not identify as users of the selected tools. 
Our interviews were centered on high level practices, which may miss low-level usage details.
Future work might follow Nam \etals{}~\cite{nam24UnderstandingLogs} logging study to understand how galleries are used in day-to-day practice rather than in recollection.
%
The libraries studied in this work have a strong influence from Heer (who is also a collaborator of ours) and some of our findings may be bound to the specific (although prominent) perspectives espoused in the communities surrounding his work.
Galleries outside of visualization may operate differently (\eg{} non-visual libraries like for data processing), and so investigating them is important future work.





\section*{Acknowledgments}
We thank our participants for sharing their time and insights with us. We also thank our reviewers for their thoughtful commentary.
This work was supported by the NSF (award \# 2141506) and the Moore Foundation.


\bibliographystyle{abbrv-doi-hyperref}
\bibliography{reference}

\begin{thebibliography}{10}

\bibitem{akbaba2023troubling}
D.~Akbaba, D.~Lange, M.~Correll, A.~Lex, and M.~Meyer.
\newblock Troubling collaboration: Matters of care for visualization design study.
\newblock In {\em SIGCHI Conference on Human Factors in Computing Systems}, pp. 1--15, 2023. \href{https://doi.org/10.1145/3544548.3581168}
{doi: {{%
10\hspace{.1pt}\discretionary{.}{%
}{.}\hspace{.4pt}1145\discretionary{/}{%
}{/}3544548\hspace{.1pt}\discretionary{.}{%
}{.}\hspace{.4pt}3581168}}}


\bibitem{bako_understanding_2022}
H.~K. Bako, X.~Liu, L.~Battle, and Z.~Liu.
\newblock Understanding {How} {Designers} {Find} and {Use} {Data} {Visualization} {Examples}.
\newblock {\em Transactions on Visualization and Computer Graphics}, pp. 1--11, 2022. \href{https://doi.org/10.1109/TVCG.2022.3209490}
{doi: {{%
10\hspace{.1pt}\discretionary{.}{%
}{.}\hspace{.4pt}1109\discretionary{/}{%
}{/}TVCG\hspace{.1pt}\discretionary{.}{%
}{.}\hspace{.4pt}2022\hspace{.1pt}\discretionary{.}{%
}{.}\hspace{.4pt}3209490}}}


\bibitem{bako_user-driven_2023}
H.~K. Bako, A.~Varma, A.~Faboro, M.~Haider, F.~Nerrise, B.~Kenah, J.~P. Dickerson, and L.~Battle.
\newblock User-driven support for visualization prototyping in d3.
\newblock In {\em International Conference on Intelligent User Interfaces}, p. 958–972. Association for Computing Machinery, New York, NY, USA, 2023. \href{https://doi.org/10.1145/3581641.3584041}
{doi: {{%
10\hspace{.1pt}\discretionary{.}{%
}{.}\hspace{.4pt}1145\discretionary{/}{%
}{/}3581641\hspace{.1pt}\discretionary{.}{%
}{.}\hspace{.4pt}3584041}}}


\bibitem{battle2022exploring}
L.~Battle, D.~Feng, and K.~Webber.
\newblock Exploring d3 implementation challenges on stack overflow.
\newblock In {\em IEEE Visualization Conference (Short Papers)}, pp. 1--5, 2022. \href{https://doi.org/10.1109/VIS54862.2022.00009}
{doi: {{%
10\hspace{.1pt}\discretionary{.}{%
}{.}\hspace{.4pt}1109\discretionary{/}{%
}{/}VIS54862\hspace{.1pt}\discretionary{.}{%
}{.}\hspace{.4pt}2022\hspace{.1pt}\discretionary{.}{%
}{.}\hspace{.4pt}00009}}}


\bibitem{bostock2011d3}
M.~Bostock, V.~Ogievetsky, and J.~Heer.
\newblock D$^3$ data-driven documents.
\newblock {\em Transactions on Visualization and Computer Graphics}, 17(12):2301--2309, 2011. \href{https://doi.org/10.1109/TVCG.2011.185}
{doi: {{%
10\hspace{.1pt}\discretionary{.}{%
}{.}\hspace{.4pt}1109\discretionary{/}{%
}{/}TVCG\hspace{.1pt}\discretionary{.}{%
}{.}\hspace{.4pt}2011\hspace{.1pt}\discretionary{.}{%
}{.}\hspace{.4pt}185}}}


\bibitem{brandt2010example}
J.~Brandt, M.~Dontcheva, M.~Weskamp, and S.~R. Klemmer.
\newblock Example-centric programming: integrating web search into the development environment.
\newblock In {\em SIGCHI Conference on Human Factors in Computing Systems}, pp. 513--522, 2010. \href{https://doi.org/10.1145/1753326.1753402}
{doi: {{%
10\hspace{.1pt}\discretionary{.}{%
}{.}\hspace{.4pt}1145\discretionary{/}{%
}{/}1753326\hspace{.1pt}\discretionary{.}{%
}{.}\hspace{.4pt}1753402}}}


\bibitem{brandt2009two}
J.~Brandt, P.~J. Guo, J.~Lewenstein, M.~Dontcheva, and S.~R. Klemmer.
\newblock Two studies of opportunistic programming: interleaving web foraging, learning, and writing code.
\newblock In {\em SIGCHI Conference on Human Factors in Computing Systems}, pp. 1589--1598, 2009. \href{https://doi.org/10.1145/1518701.1518944}
{doi: {{%
10\hspace{.1pt}\discretionary{.}{%
}{.}\hspace{.4pt}1145\discretionary{/}{%
}{/}1518701\hspace{.1pt}\discretionary{.}{%
}{.}\hspace{.4pt}1518944}}}


\bibitem{conlen2018idyll}
M.~Conlen and J.~Heer.
\newblock Idyll: A markup language for authoring and publishing interactive articles on the web.
\newblock In {\em ACM Symposium on User Interface Software and Technology}, pp. 977--989, 2018. \href{https://doi.org/10.1145/3242587.3242600}
{doi: {{%
10\hspace{.1pt}\discretionary{.}{%
}{.}\hspace{.4pt}1145\discretionary{/}{%
}{/}3242587\hspace{.1pt}\discretionary{.}{%
}{.}\hspace{.4pt}3242600}}}


\bibitem{correll2019ethical}
M.~Correll.
\newblock Ethical dimensions of visualization research.
\newblock In {\em SIGCHI conference on human factors in computing systems}, pp. 1--13, 2019. \href{https://doi.org/10.1145/3290605.3300418}
{doi: {{%
10\hspace{.1pt}\discretionary{.}{%
}{.}\hspace{.4pt}1145\discretionary{/}{%
}{/}3290605\hspace{.1pt}\discretionary{.}{%
}{.}\hspace{.4pt}3300418}}}


\bibitem{crichton2020documentation}
W.~Crichton.
\newblock Documentation generation as information visualization.
\newblock {\em PLATEAU Annual Workshop on the Intersection of HCI and PL}, 2020. \href{https://arxiv.org/abs/2011.05600}
{doi: {{%
abs\discretionary{/}{%
}{/}2011\hspace{.1pt}\discretionary{.}{%
}{.}\hspace{.4pt}05600}}}


\bibitem{crichton21Nota}
W.~Crichton.
\newblock A new medium for communicating research on programming languages.
\newblock In {\em Workshop on Human Aspects of Types and Reasoning Assistants}, 2021.

\bibitem{daka2014survey}
E.~Daka and G.~Fraser.
\newblock A survey on unit testing practices and problems.
\newblock In {\em International Symposium on Software Reliability Engineering}, pp. 201--211, 2014. \href{https://doi.org/10.1109/ISSRE.2014.11}
{doi: {{%
10\hspace{.1pt}\discretionary{.}{%
}{.}\hspace{.4pt}1109\discretionary{/}{%
}{/}ISSRE\hspace{.1pt}\discretionary{.}{%
}{.}\hspace{.4pt}2014\hspace{.1pt}\discretionary{.}{%
}{.}\hspace{.4pt}11}}}


\bibitem{sphinx}
T.~S. developers.
\newblock Sphinx.
\newblock \url{https://www.sphinx-doc.org/}, 2007.

\bibitem{fowler2010domain}
M.~Fowler.
\newblock {\em Domain-specific languages}.
\newblock Pearson Education, 2010.

\bibitem{geiger_types_2018}
R.~S. Geiger, N.~Varoquaux, C.~Mazel-Cabasse, and C.~Holdgraf.
\newblock The {Types}, {Roles}, and {Practices} of {Documentation} in {Data} {Analytics} {Open} {Source} {Software} {Libraries}: {A} {Collaborative} {Ethnography} of {Documentation} {Work}.
\newblock {\em Computer Supported Cooperative Work}, 27(3-6):767--802, 2018. \href{https://doi.org/10.1007/s10606-018-9333-1}
{doi: {{%
10\hspace{.1pt}\discretionary{.}{%
}{.}\hspace{.4pt}1007\discretionary{/}{%
}{/}s10606\discretionary{%
}{-}{-}018\discretionary{%
}{-}{-}9333\discretionary{%
}{-}{-}1}}}


\bibitem{glassman_visualizing_2018}
E.~L. Glassman, T.~Zhang, B.~Hartmann, and M.~Kim.
\newblock Visualizing {API} {Usage} {Examples} at {Scale}.
\newblock In {\em SIGCHI Conference on Human Factors in Computing Systems}, pp. 1--12. ACM, 2018. \href{https://doi.org/10.1145/3173574.3174154}
{doi: {{%
10\hspace{.1pt}\discretionary{.}{%
}{.}\hspace{.4pt}1145\discretionary{/}{%
}{/}3173574\hspace{.1pt}\discretionary{.}{%
}{.}\hspace{.4pt}3174154}}}


\bibitem{head2015tutorons}
A.~Head, C.~Appachu, M.~A. Hearst, and B.~Hartmann.
\newblock Tutorons: Generating context-relevant, on-demand explanations and demonstrations of online code.
\newblock In {\em Symposium on Visual Languages and Human-Centric Computing}, pp. 3--12. IEEE, 2015. \href{https://doi.org/10.1109/VLHCC.2015.7356972}
{doi: {{%
10\hspace{.1pt}\discretionary{.}{%
}{.}\hspace{.4pt}1109\discretionary{/}{%
}{/}VLHCC\hspace{.1pt}\discretionary{.}{%
}{.}\hspace{.4pt}2015\hspace{.1pt}\discretionary{.}{%
}{.}\hspace{.4pt}7356972}}}


\bibitem{head_composing_2020}
A.~Head, J.~Jiang, J.~Smith, M.~A. Hearst, and B.~Hartmann.
\newblock Composing {Flexibly}-{Organized} {Step}-by-{Step} {Tutorials} from {Linked} {Source} {Code}, {Snippets}, and {Outputs}.
\newblock In {\em SIGCHI Conference on Human Factors in Computing Systems}, pp. 1--12. ACM, 2020. \href{https://doi.org/10.1145/3313831.3376798}
{doi: {{%
10\hspace{.1pt}\discretionary{.}{%
}{.}\hspace{.4pt}1145\discretionary{/}{%
}{/}3313831\hspace{.1pt}\discretionary{.}{%
}{.}\hspace{.4pt}3376798}}}


\bibitem{hengesbach2022undoing}
N.~Hengesbach.
\newblock Undoing seamlessness: Exploring seams for critical visualization.
\newblock In {\em CHI Conference on Human Factors in Computing Systems Extended Abstracts}, pp. 1--7, 2022. \href{https://doi.org/10.1145/3491101.3519703}
{doi: {{%
10\hspace{.1pt}\discretionary{.}{%
}{.}\hspace{.4pt}1145\discretionary{/}{%
}{/}3491101\hspace{.1pt}\discretionary{.}{%
}{.}\hspace{.4pt}3519703}}}


\bibitem{hoffswell2016visual}
J.~Hoffswell, A.~Satyanarayan, and J.~Heer.
\newblock Visual debugging techniques for reactive data visualization.
\newblock In {\em Computer Graphics Forum}, vol.~35, pp. 271--280. Wiley, 2016. \href{https://doi.org/10.1111/cgf.12903}
{doi: {{%
10\hspace{.1pt}\discretionary{.}{%
}{.}\hspace{.4pt}1111\discretionary{/}{%
}{/}cgf\hspace{.1pt}\discretionary{.}{%
}{.}\hspace{.4pt}12903}}}


\bibitem{horvath2022understanding}
A.~Horvath, M.~X. Liu, R.~Hendriksen, C.~Shannon, E.~Paterson, K.~Jawad, A.~Macvean, and B.~A. Myers.
\newblock Understanding how programmers can use annotations on documentation.
\newblock In {\em SIGCHI Conference on Human Factors in Computing Systems}, pp. 1--16, 2022. \href{https://doi.org/10.1145/3491102.3502095}
{doi: {{%
10\hspace{.1pt}\discretionary{.}{%
}{.}\hspace{.4pt}1145\discretionary{/}{%
}{/}3491102\hspace{.1pt}\discretionary{.}{%
}{.}\hspace{.4pt}3502095}}}


\bibitem{horvath2023support}
A.~Horvath, A.~Macvean, and B.~A. Myers.
\newblock Support for long-form documentation authoring and maintenance.
\newblock In {\em Symposium on Visual Languages and Human-Centric Computing}, pp. 109--114. IEEE, 2023. \href{https://doi.org/10.1109/VL-HCC57772.2023.00020}
{doi: {{%
10\hspace{.1pt}\discretionary{.}{%
}{.}\hspace{.4pt}1109\discretionary{/}{%
}{/}VL\discretionary{%
}{-}{-}HCC57772\hspace{.1pt}\discretionary{.}{%
}{.}\hspace{.4pt}2023\hspace{.1pt}\discretionary{.}{%
}{.}\hspace{.4pt}00020}}}


\bibitem{hu2019viznet}
K.~Hu, S.~Gaikwad, M.~Hulsebos, M.~A. Bakker, E.~Zgraggen, C.~Hidalgo, T.~Kraska, G.~Li, A.~Satyanarayan, and {\c{C}}.~Demiralp.
\newblock Viznet: Towards a large-scale visualization learning and benchmarking repository.
\newblock In {\em SIGCHI conference on human factors in computing systems}, pp. 1--12, 2019. \href{https://doi.org/10.1145/3290605.3300892}
{doi: {{%
10\hspace{.1pt}\discretionary{.}{%
}{.}\hspace{.4pt}1145\discretionary{/}{%
}{/}3290605\hspace{.1pt}\discretionary{.}{%
}{.}\hspace{.4pt}3300892}}}


\bibitem{ichinco_exploring_2015}
M.~Ichinco and C.~Kelleher.
\newblock Exploring novice programmer example use.
\newblock In {\em Symposium on Visual Languages and Human-Centric Computing}, pp. 63--71. IEEE, 2015. \href{https://doi.org/10.1109/VLHCC.2015.7357199}
{doi: {{%
10\hspace{.1pt}\discretionary{.}{%
}{.}\hspace{.4pt}1109\discretionary{/}{%
}{/}VLHCC\hspace{.1pt}\discretionary{.}{%
}{.}\hspace{.4pt}2015\hspace{.1pt}\discretionary{.}{%
}{.}\hspace{.4pt}7357199}}}


\bibitem{jakubovic2023technical}
J.~Jakubovic, J.~Edwards, and T.~Petricek.
\newblock {Technical Dimensions of Programming Systems}.
\newblock {\em Art, Science, and Engineering of Programming}, 7(3), 2023. \href{https://doi.org/10.22152/programming-journal.org/2023/7/13}
{doi: {{%
10\hspace{.1pt}\discretionary{.}{%
}{.}\hspace{.4pt}22152\discretionary{/}{%
}{/}programming\discretionary{%
}{-}{-}journal\hspace{.1pt}\discretionary{.}{%
}{.}\hspace{.4pt}org\discretionary{/}{%
}{/}2023\discretionary{/}{%
}{/}7\discretionary{/}{%
}{/}13}}}


\bibitem{jernigan_principled_2015}
W.~Jernigan, A.~Horvath, M.~Lee, M.~Burnett, T.~Cuilty, S.~Kuttal, A.~Peters, I.~Kwan, F.~Bahmani, and A.~Ko.
\newblock A principled evaluation for a principled idea garden.
\newblock In {\em Symposium on Visual Languages and Human-Centric Computing}, pp. 235--243. IEEE, 2015. \href{https://doi.org/10.1109/VLHCC.2015.7357222}
{doi: {{%
10\hspace{.1pt}\discretionary{.}{%
}{.}\hspace{.4pt}1109\discretionary{/}{%
}{/}VLHCC\hspace{.1pt}\discretionary{.}{%
}{.}\hspace{.4pt}2015\hspace{.1pt}\discretionary{.}{%
}{.}\hspace{.4pt}7357222}}}


\bibitem{kang_paragon_2018}
H.~B. Kang, G.~Amoako, N.~Sengupta, and S.~P. Dow.
\newblock Paragon: {An} {Online} {Gallery} for {Enhancing} {Design} {Feedback} with {Visual} {Examples}.
\newblock In {\em SIGCHI Conference on Human Factors in Computing Systems}, pp. 1--13. ACM, 2018. \href{https://doi.org/10.1145/3173574.3174180}
{doi: {{%
10\hspace{.1pt}\discretionary{.}{%
}{.}\hspace{.4pt}1145\discretionary{/}{%
}{/}3173574\hspace{.1pt}\discretionary{.}{%
}{.}\hspace{.4pt}3174180}}}


\bibitem{kruchten_metrics-based_2023}
N.~Kruchten, A.~M. McNutt, and M.~J. McGuffin.
\newblock {Metrics-Based Evaluation and Comparison of Visualization Notations}.
\newblock {\em IEEE Transactions on Visualization and Computer Graphics}, 2023. \href{https://doi.org/10.1109/TVCG.2023.3326907}
{doi: {{%
10\hspace{.1pt}\discretionary{.}{%
}{.}\hspace{.4pt}1109\discretionary{/}{%
}{/}TVCG\hspace{.1pt}\discretionary{.}{%
}{.}\hspace{.4pt}2023\hspace{.1pt}\discretionary{.}{%
}{.}\hspace{.4pt}3326907}}}


\bibitem{le_moulec_automatic_2018}
G.~Le~Moulec, A.~Blouin, V.~Gouranton, and B.~Arnaldi.
\newblock Automatic production of end user documentation for {DSLs}.
\newblock {\em Computer Languages, Systems \& Structures}, 54:337--357, 2018. \href{https://doi.org/10.1016/j.cl.2018.07.006}
{doi: {{%
10\hspace{.1pt}\discretionary{.}{%
}{.}\hspace{.4pt}1016\discretionary{/}{%
}{/}j\hspace{.1pt}\discretionary{.}{%
}{.}\hspace{.4pt}cl\hspace{.1pt}\discretionary{.}{%
}{.}\hspace{.4pt}2018\hspace{.1pt}\discretionary{.}{%
}{.}\hspace{.4pt}07\hspace{.1pt}\discretionary{.}{%
}{.}\hspace{.4pt}006}}}


\bibitem{lee_designing_2010}
B.~Lee, S.~Srivastava, R.~Kumar, R.~Brafman, and S.~R. Klemmer.
\newblock Designing with interactive example galleries.
\newblock In {\em SIGCHI Conference on Human Factors in Computing Systems}, pp. 2257--2266. ACM, 2010. \href{https://doi.org/10.1145/1753326.1753667}
{doi: {{%
10\hspace{.1pt}\discretionary{.}{%
}{.}\hspace{.4pt}1145\discretionary{/}{%
}{/}1753326\hspace{.1pt}\discretionary{.}{%
}{.}\hspace{.4pt}1753667}}}


\bibitem{lee2022affective}
E.~Lee-Robbins and E.~Adar.
\newblock Affective learning objectives for communicative visualizations.
\newblock {\em IEEE Transactions on Visualization and Computer Graphics}, 29(1):1--11, 2022. \href{https://doi.org/10.1109/TVCG.2022.3209500}
{doi: {{%
10\hspace{.1pt}\discretionary{.}{%
}{.}\hspace{.4pt}1109\discretionary{/}{%
}{/}TVCG\hspace{.1pt}\discretionary{.}{%
}{.}\hspace{.4pt}2022\hspace{.1pt}\discretionary{.}{%
}{.}\hspace{.4pt}3209500}}}


\bibitem{luceroAffinity}
A.~Lucero.
\newblock Using affinity diagrams to evaluate interactive prototypes.
\newblock In {\em Human-Computer Interaction - INTERACT}, vol. 9297 of {\em Lecture Notes in Computer Science}, pp. 231--248. Springer, 2015. \href{https://doi.org/10.1007/978-3-319-22668-2_19}
{doi: {{%
10\hspace{.1pt}\discretionary{.}{%
}{.}\hspace{.4pt}1007\discretionary{/}{%
}{/}978\discretionary{%
}{-}{-}3\discretionary{%
}{-}{-}319\discretionary{%
}{-}{-}22668\discretionary{%
}{-}{-}2\_19}}}


\bibitem{information_is_beautiful}
D.~McCandless.
\newblock Information is beautiful.
\newblock Accessed March 25, 2024.

\bibitem{mcnutt_integrated_2021}
A.~M. McNutt and R.~Chugh.
\newblock Integrated visualization editing via parameterized declarative templates.
\newblock In {\em SIGCHI Conference on Human Factors in Computing Systems}. ACM, NYC, 2021. \href{https://doi.org/10.1145/3411764.3445356}
{doi: {{%
10\hspace{.1pt}\discretionary{.}{%
}{.}\hspace{.4pt}1145\discretionary{/}{%
}{/}3411764\hspace{.1pt}\discretionary{.}{%
}{.}\hspace{.4pt}3445356}}}


\bibitem{mcnutt_study_2023}
A.~M. McNutt, A.~Outkine, and R.~Chugh.
\newblock {A Study of Editor Features in a Creative Coding Classroom}.
\newblock In {\em SIGCHI Conference on Human Factors in Computing Systems}. Association for Computing Machinery, NYC, 2023. \href{https://doi.org/10.1145/3544548.3580683}
{doi: {{%
10\hspace{.1pt}\discretionary{.}{%
}{.}\hspace{.4pt}1145\discretionary{/}{%
}{/}3544548\hspace{.1pt}\discretionary{.}{%
}{.}\hspace{.4pt}3580683}}}


\bibitem{mehrpour_active_2019}
S.~Mehrpour, T.~D. LaToza, and R.~K. Kindi.
\newblock Active {Documentation}: {Helping} {Developers} {Follow} {Design} {Decisions}.
\newblock In {\em Symposium on Visual Languages and Human-Centric Computing}, pp. 87--96. IEEE, 2019. \href{https://doi.org/10.1109/VLHCC.2019.8818816}
{doi: {{%
10\hspace{.1pt}\discretionary{.}{%
}{.}\hspace{.4pt}1109\discretionary{/}{%
}{/}VLHCC\hspace{.1pt}\discretionary{.}{%
}{.}\hspace{.4pt}2019\hspace{.1pt}\discretionary{.}{%
}{.}\hspace{.4pt}8818816}}}


\bibitem{mooty2010calcite}
M.~Mooty, A.~Faulring, J.~Stylos, and B.~A. Myers.
\newblock Calcite: Completing code completion for constructors using crowds.
\newblock In {\em 2010 IEEE Symposium on Visual Languages and Human-Centric Computing}, pp. 15--22. IEEE, 2010. \href{https://doi.org/10.1109/VLHCC.2010.12}
{doi: {{%
10\hspace{.1pt}\discretionary{.}{%
}{.}\hspace{.4pt}1109\discretionary{/}{%
}{/}VLHCC\hspace{.1pt}\discretionary{.}{%
}{.}\hspace{.4pt}2010\hspace{.1pt}\discretionary{.}{%
}{.}\hspace{.4pt}12}}}


\bibitem{nam24UnderstandingLogs}
D.~Nam, A.~Macvean, B.~Myers, and B.~Vasilescu.
\newblock Understanding documentation use through log analysis an exploratory case study of four cloud services.
\newblock {\em SIGCHI Conference on Human Factors in Computing Systems}, 2024.
\newblock To Appear. \href{https://doi.org/10.48550/arXiv.2310.10817}
{doi: {{%
10\hspace{.1pt}\discretionary{.}{%
}{.}\hspace{.4pt}48550\discretionary{/}{%
}{/}arXiv\hspace{.1pt}\discretionary{.}{%
}{.}\hspace{.4pt}2310\hspace{.1pt}\discretionary{.}{%
}{.}\hspace{.4pt}10817}}}


\bibitem{narechania2020nl4dv}
A.~Narechania, A.~Srinivasan, and J.~Stasko.
\newblock Nl4dv: A toolkit for generating analytic specifications for data visualization from natural language queries.
\newblock {\em Transactions on Visualization and Computer Graphics}, 27(2):369--379, 2020. \href{https://doi.org/10.1109/TVCG.2020.3030378}
{doi: {{%
10\hspace{.1pt}\discretionary{.}{%
}{.}\hspace{.4pt}1109\discretionary{/}{%
}{/}TVCG\hspace{.1pt}\discretionary{.}{%
}{.}\hspace{.4pt}2020\hspace{.1pt}\discretionary{.}{%
}{.}\hspace{.4pt}3030378}}}


\bibitem{nasehi2010unit}
S.~M. Nasehi and F.~Maurer.
\newblock Unit tests as api usage examples.
\newblock In {\em 2013 IEEE International Conference on Software Maintenance}, pp. 1--10. IEEE Computer Society, Los Alamitos, CA, USA, sep 2010. \href{https://doi.org/10.1109/ICSM.2010.5609553}
{doi: {{%
10\hspace{.1pt}\discretionary{.}{%
}{.}\hspace{.4pt}1109\discretionary{/}{%
}{/}ICSM\hspace{.1pt}\discretionary{.}{%
}{.}\hspace{.4pt}2010\hspace{.1pt}\discretionary{.}{%
}{.}\hspace{.4pt}5609553}}}


\bibitem{nasehi_what_2012}
S.~M. Nasehi, J.~Sillito, F.~Maurer, and C.~Burns.
\newblock What makes a good code example?: {A} study of programming {Q}\&{A} in {StackOverflow}.
\newblock In {\em International Conference on Software Maintenance}, pp. 25--34. IEEE, 2012. \href{https://doi.org/10.1109/ICSM.2012.6405249}
{doi: {{%
10\hspace{.1pt}\discretionary{.}{%
}{.}\hspace{.4pt}1109\discretionary{/}{%
}{/}ICSM\hspace{.1pt}\discretionary{.}{%
}{.}\hspace{.4pt}2012\hspace{.1pt}\discretionary{.}{%
}{.}\hspace{.4pt}6405249}}}


\bibitem{deckgl}
{OpenJS Foundation}.
\newblock deck.gl.
\newblock \url{https://deck.gl/}.
\newblock Accessed March 25, 2024.

\bibitem{oppermann_vizsnippets_2022}
M.~Oppermann and T.~Munzner.
\newblock {VizSnippets}: {Compressing} {Visualization} {Bundles} {Into} {Representative} {Previews} for {Browsing} {Visualization} {Collections}.
\newblock {\em Transactions on Visualization and Computer Graphics}, 28(1):747--757, 2022. \href{https://doi.org/10.1109/TVCG.2021.3114841}
{doi: {{%
10\hspace{.1pt}\discretionary{.}{%
}{.}\hspace{.4pt}1109\discretionary{/}{%
}{/}TVCG\hspace{.1pt}\discretionary{.}{%
}{.}\hspace{.4pt}2021\hspace{.1pt}\discretionary{.}{%
}{.}\hspace{.4pt}3114841}}}


\bibitem{p5js}
p5.js.
\newblock examples.
\newblock \url{https://p5js.org/examples/}.
\newblock Accessed March 25, 2024.

\bibitem{plotly}
Plotly.
\newblock Low-code python data app.
\newblock \url{https://plotly.com/}.
\newblock Accessed March 25, 2024.

\bibitem{quarto}
Posit.
\newblock Quarto.
\newblock \url{https://quarto.org/}.
\newblock Accessed March 25, 2024.

\bibitem{potter2022contextualized}
H.~Potter, A.~Madadi, R.~Just, and C.~Omar.
\newblock Contextualized programming language documentation.
\newblock In {\em ACM SIGPLAN International Symposium on New Ideas, New Paradigms, and Reflections on Programming and Software}, pp. 1--15, 2022. \href{https://doi.org/10.1145/3563835.3567654}
{doi: {{%
10\hspace{.1pt}\discretionary{.}{%
}{.}\hspace{.4pt}1145\discretionary{/}{%
}{/}3563835\hspace{.1pt}\discretionary{.}{%
}{.}\hspace{.4pt}3567654}}}


\bibitem{ProcidaDiataxis}
D.~Procida.
\newblock Diátaxis: A systematic approach to technical documentation authoring.
\newblock \url{https://diataxis.fr/#}.
\newblock Accessed March 29, 2024.

\bibitem{reas2007processing}
C.~Reas and B.~Fry.
\newblock {\em Processing: a programming handbook for visual designers and artists}, vol. 6812.
\newblock Mit Press, 2007. \href{https://www.worldcat.org/oclc/73993935}
{doi: {{%
oclc\discretionary{/}{%
}{/}73993935}}}


\bibitem{sabab2020automated}
S.~A. Sabab, A.~Khan, P.~K. Chilana, J.~McGrenere, and A.~Bunt.
\newblock An automated approach to assessing an application tutorial’s difficulty.
\newblock In {\em 2020 IEEE Symposium on Visual Languages and Human-Centric Computing}, pp. 1--10. IEEE, 2020. \href{https://doi.org/10.1109/VL/HCC50065.2020.9127271}
{doi: {{%
10\hspace{.1pt}\discretionary{.}{%
}{.}\hspace{.4pt}1109\discretionary{/}{%
}{/}VL\discretionary{/}{%
}{/}HCC50065\hspace{.1pt}\discretionary{.}{%
}{.}\hspace{.4pt}2020\hspace{.1pt}\discretionary{.}{%
}{.}\hspace{.4pt}9127271}}}


\bibitem{satyanarayan2016vega}
A.~Satyanarayan, D.~Moritz, K.~Wongsuphasawat, and J.~Heer.
\newblock Vega-lite: A grammar of interactive graphics.
\newblock {\em Transactions on Visualization and Computer Graphics}, 23(1):341--350, 2016. \href{https://doi.org/10.1109/TVCG.2016.2599030}
{doi: {{%
10\hspace{.1pt}\discretionary{.}{%
}{.}\hspace{.4pt}1109\discretionary{/}{%
}{/}TVCG\hspace{.1pt}\discretionary{.}{%
}{.}\hspace{.4pt}2016\hspace{.1pt}\discretionary{.}{%
}{.}\hspace{.4pt}2599030}}}


\bibitem{satyanarayan2015reactive}
A.~Satyanarayan, R.~Russell, J.~Hoffswell, and J.~Heer.
\newblock Reactive vega: A streaming dataflow architecture for declarative interactive visualization.
\newblock {\em Transactions on Visualization and Computer Graphics}, 22(1):659--668, 2015. \href{https://doi.org/10.1109/TVCG.2015.2467091}
{doi: {{%
10\hspace{.1pt}\discretionary{.}{%
}{.}\hspace{.4pt}1109\discretionary{/}{%
}{/}TVCG\hspace{.1pt}\discretionary{.}{%
}{.}\hspace{.4pt}2015\hspace{.1pt}\discretionary{.}{%
}{.}\hspace{.4pt}2467091}}}


\bibitem{setlur_olio_2023}
V.~Setlur, A.~Kanyuka, and A.~Srinivasan.
\newblock Olio: {A} {Semantic} {Search} {Interface} for {Data} {Repositories}.
\newblock In {\em ACM Symposium on User Interface Software and Technology}, pp. 1--16. ACM, 2023. \href{https://doi.org/10.1145/3586183.3606806}
{doi: {{%
10\hspace{.1pt}\discretionary{.}{%
}{.}\hspace{.4pt}1145\discretionary{/}{%
}{/}3586183\hspace{.1pt}\discretionary{.}{%
}{.}\hspace{.4pt}3606806}}}


\bibitem{subbaraman2023forking}
B.~Subbaraman, S.~Shim, and N.~Peek.
\newblock Forking a sketch: How the openprocessing community uses remixing to collect, annotate, tune, and extend creative code.
\newblock In {\em ACM Designing Interactive Systems Conference}, pp. 326--342, 2023. \href{https://doi.org/10.1145/3563657.3595969}
{doi: {{%
10\hspace{.1pt}\discretionary{.}{%
}{.}\hspace{.4pt}1145\discretionary{/}{%
}{/}3563657\hspace{.1pt}\discretionary{.}{%
}{.}\hspace{.4pt}3595969}}}


\bibitem{tanzil2024people}
M.~H. Tanzil, G.~Uddin, and A.~Barcomb.
\newblock " how do people decide?": A model for software library selection.
\newblock In {\em IEEE/ACM International Conference on Cooperative and Human Aspects of Software Engineering}, 2024. \href{https://doi.org/10.48550/arXiv.2403.16245}
{doi: {{%
10\hspace{.1pt}\discretionary{.}{%
}{.}\hspace{.4pt}48550\discretionary{/}{%
}{/}arXiv\hspace{.1pt}\discretionary{.}{%
}{.}\hspace{.4pt}2403\hspace{.1pt}\discretionary{.}{%
}{.}\hspace{.4pt}16245}}}


\bibitem{thayer_theory_2021}
K.~Thayer, S.~E. Chasins, and A.~J. Ko.
\newblock A {Theory} of {Robust} {API} {Knowledge}.
\newblock {\em ACM Transactions on Computing Education}, 21(1):1--32, 2021. \href{https://doi.org/10.1145/3444945}
{doi: {{%
10\hspace{.1pt}\discretionary{.}{%
}{.}\hspace{.4pt}1145\discretionary{/}{%
}{/}3444945}}}


\bibitem{vega-docs}
Vega.
\newblock Documentation.
\newblock \url{https://vega.github.io/vega/docs}.
\newblock Accessed March 29, 2024.

\bibitem{vega-internal}
Vega.
\newblock Documentation.
\newblock \url{https://observablehq.com/@vega/how-vega-works}.
\newblock Accessed March 29, 2024.

\bibitem{vega-editor}
Vega.
\newblock Editor.
\newblock \url{https://vega.github.io/editor/}.
\newblock Accessed March 29, 2024.

\bibitem{vega-bar-chart}
Vega.
\newblock Let's make a bar chart tutorial.
\newblock \url{https://vega.github.io/vega/tutorials/bar-chart/}.
\newblock Accessed March 29, 2024.

\bibitem{ipyvizzu}
{Vizzu Inc.}
\newblock ipyvizzu.
\newblock \url{https://ipyvizzu.vizzuhq.com/latest/}.
\newblock Accessed March 25, 2024.

\bibitem{wang_novices_2021}
W.~Wang, A.~Kwatra, J.~Skripchuk, N.~Gomes, A.~Milliken, C.~Martens, T.~Barnes, and T.~Price.
\newblock Novices' {Learning} {Barriers} {When} {Using} {Code} {Examples} in {Open}-{Ended} {Programming}.
\newblock {\em ACM Conference on Innovation and Technology in Computer Science Education}, pp. 394--400, 2021.
\newblock ACM Conference on Innovation and Technology in Computer Science Education ISBN: 9781450382144 Place: Virtual Event Germany Publisher: ACM. \href{https://doi.org/10.1145/3430665.3456370}
{doi: {{%
10\hspace{.1pt}\discretionary{.}{%
}{.}\hspace{.4pt}1145\discretionary{/}{%
}{/}3430665\hspace{.1pt}\discretionary{.}{%
}{.}\hspace{.4pt}3456370}}}


\bibitem{xu_ideaterelate_2021}
X.~T. Xu, R.~Xiong, B.~Wang, D.~Min, and S.~P. Dow.
\newblock {IdeateRelate}: {An} {Examples} {Gallery} {That} {Helps} {Creators} {Explore} {Ideas} in {Relation} to {Their} {Own}.
\newblock {\em Proceedings of the ACM on Human-Computer Interaction}, 5:1--18, 2021. \href{https://doi.org/10.1145/3479496}
{doi: {{%
10\hspace{.1pt}\discretionary{.}{%
}{.}\hspace{.4pt}1145\discretionary{/}{%
}{/}3479496}}}


\bibitem{yang2023vegaprof}
J.~Yang, A.~B{\"a}uerle, D.~Moritz, and {\c{C}}.~Demiralp.
\newblock Vegaprof: Profiling vega visualizations.
\newblock In {\em ACM Symposium on User Interface Software and Technology}, pp. 1--11, 2023. \href{https://doi.org/10.1145/3586183.3606790}
{doi: {{%
10\hspace{.1pt}\discretionary{.}{%
}{.}\hspace{.4pt}1145\discretionary{/}{%
}{/}3586183\hspace{.1pt}\discretionary{.}{%
}{.}\hspace{.4pt}3606790}}}


\bibitem{yang2023draco2}
J.~Yang, P.~F. Gyarmati, Z.~Zeng, and D.~Moritz.
\newblock Draco 2: An extensible platform to model visualization design.
\newblock In {\em IEEE Visualization and Visual Analytics}, pp. 166--170, 2023. \href{https://doi.org/10.1109/VIS54172.2023.00042}
{doi: {{%
10\hspace{.1pt}\discretionary{.}{%
}{.}\hspace{.4pt}1109\discretionary{/}{%
}{/}VIS54172\hspace{.1pt}\discretionary{.}{%
}{.}\hspace{.4pt}2023\hspace{.1pt}\discretionary{.}{%
}{.}\hspace{.4pt}00042}}}


\bibitem{ye2020penrose}
K.~Ye, W.~Ni, M.~Krieger, D.~Ma'ayan, J.~Wise, J.~Aldrich, J.~Sunshine, and K.~Crane.
\newblock Penrose: from mathematical notation to beautiful diagrams.
\newblock {\em ACM Transactions on Graphics}, 39(4):144--1, 2020. \href{https://doi.org/10.1145/3386569.3392375}
{doi: {{%
10\hspace{.1pt}\discretionary{.}{%
}{.}\hspace{.4pt}1145\discretionary{/}{%
}{/}3386569\hspace{.1pt}\discretionary{.}{%
}{.}\hspace{.4pt}3392375}}}


\end{thebibliography}

\clearpage
\appendix{}

\section{Interview Guides}

In this appendix we reproduced the interview guides used in our semi-structured interviews. Follow up questions are collapsed into single bullet items to simplify the presentation. 

\revised{
\subsection{Creator Backgrounds}

Here we review in greater detail the relationships that the authors of this work have with our participants as a means to better concretize our positionality.
The three creators, Jeffrey Heer, Dominik Moritz, and Matt Conlen are, or have been, affiliated with the same research lab (UW IDL) as all three of the authors of this work. 
Carlos Scheidegger, Jeffrey Heer, and Dominik Moritz are or have been current collaborators of Leilani’s. 
Andrew co-authored a paper with Nicolas Kruchten, worked with Xiaoji Chen and Ib Green, and was advised by Jeffrey Heer. 
Junran has co-authored papers with Dominik Moritz and Jeffrey Heer. 
This work focuses soliciting the experiences of a small subject population who might potentially share common preferences and habits for gallery designing from previous collaboration experience; future research engaging with tool builders more exclusively is likely to discover additional insights in gallery creation. 
This work also focused on users of those galleries, but there was no overlap between the research team and that group.
}

\subsection{Gallery Creator Interview Guide}

\parahead{Study introduction}
Hi, \emph{name}! Thank you for participating in today's interview and volunteering your time to help us understand example gallery design.

\emph{Interviewers introduce themselves (name, position). }

To quickly re-summarize our work, we are investigating why and how DSL developers create example galleries as a part of their documentation. In this interview, you will share the design decisions and the process for creating the galleries. Feel free to share your screen at any time to answer our questions.

If you don't have any questions and you consent to participating I will begin to record this meeting, is that okay with you? If yes, start the recording.

\parahead{Opening questions - background and introduction}

\begin{itemize}
     \item I'm familiar with [the DSL], but can you tell me about how you view it and your role in developing it? What makes it different from existing tools
     \item Who are the target groups and how do you expect them to start using it?
     \item Like domain expert or practiced programmer, one-time user or frequent user, for job or for fun? Have you noticed any unexpected user groups?
     \item In your library and documentation, what things are considered as galleries? What is the purpose of each of them? Like advertisement, learning resources, direct reuse, test, standard; for users or for your own
\end{itemize}

\parahead{Documentation and gallery}

\begin{itemize}
     \item Besides galleries, what resources do you include? For example, tutorial, blog post, detailed API docs, and code examples What's the most important format of resources? How are these different pieces connected? How should a user work with these resources to use your library efficiently?
     \item How is a gallery different from all of the other documentation formats? Can you give some examples for how they complement the others?
     \item Are there tools you use to create and manage the documentation? Could you describe your experience using these tools? Which part of the documentation do you use them for? Are you satisfied with these tools and what features do you think would be great to have in an ideal tool?
\end{itemize}

\parahead{Gallery design and maintenance}

\begin{itemize}
     \item If you were going to show only a few examples (from your DSL gallery) to someone, which example would it be? Which are the top or bottom well-designed (pick one)? What about them do you like or hate? For first-time users, which would you suggest? How do you think they would benefit? Would you suggest any examples to programming novices but domain users? How do you think they would benefit from them?
     \item  Tell us about the process of designing a new example? Do you make them manually or repurpose existing ones? Are there galleries that inspired you and can you explain the connection with your gallery examples?
     \item  From where did you learn to do gallery design? Are there papers, blog posts or books you used? Galleries other than the previous ones we talked about? Do you find them useful? Do you disagree with any of the information? Examples? What would you suggest to someone with less experience?

     \item What is the process of creating the gallery? Can you describe the timeline? When do you start to work on documentation and galleries? When do you consider the gallery mostly done?

     \item What are some key decisions you have to make for design and modification? What do you need to help make these decisions?

     \item How do you figure out how many examples to include? What's the largest gallery you've created? And why that many? Would you include as many examples as possible? Are there any organizational challenges?

     \item If you were to write down your own “lessons learned” or “rules of thumb” or “best practices” about gallery design, what would it include? E.g. Maximize the coverage of features; From simple to complex examples; categorize. How did you learn these?

     \item Is there any connection between the gallery and your test suite?
     \item How do you maintain the examples? When do you add, update or delete examples? How do you decide whether or not to include a certain example? Can you give some examples? When do you need to sort or categorize the examples? How do you order the categories? Do you put the same examples under multiple categories? If not, how do you choose which category to put when there are multiple options?
     \item  Are there additional features you would provide to support searching and browsing the examples? E.g. meaningful names, keywords, keyword search, categories, links to documentations, link to related examples
     \item given infinite resources (time/money) what would you want your gallery to look like?
     \item What do you care about when you evaluate the gallery? Like usability, learnability, quality. Any metrics? Does this differ based on the library itself?
     \item Were there any challenges you had creating or maintaining the galleries?
     \item Any hidden assumptions, usage patterns that are difficult to perceive by the gallery reader, compared to the visual things? How do you incorporate them in the docs

\end{itemize}

\parahead{User feedbacks}

\begin{itemize}
     \item Do your users have a communication channel back to you or the other developers? for instance, GitHub issue, a survey you all did, etc. What do they request about docs and galleries? Has this been helpful? Do you modify anything based on their feedback? Examples? Do you think they understand why you do things in the way you do? Have you seen any complaints about the gallery examples? Have you seen examples being misused?
     \item Are there any other kinds of data that you collect from users? Like user surveys?
     \item In general, how much do you think the gallery affects the adoption of your library?
\end{itemize}

You can answer these question either now or in the exit survey:

\begin{itemize}
     \item Any other thoughts on galleries that we haven't covered?
     \item Would you recommend anyone as our interviewee?
\end{itemize}
Thank you very much! These have been very useful!


\subsection{Gallery User Interview Guide}

Intro as we talked about during the meeting “Today we're interested in talking about the galleries for vl/altair/plotly. Which of these are you most familiar with and how long have you been using it 

The goals of the interviews are to understand how people use example galleries (to learn or to accomplish their tasks efficiently), and to understand opportunities for better galleries and tools to help people learn and create visualizations.  

\begin{itemize}

\item What made you start to use [DSL]? For what tasks? Why did you choose it over other alternatives? 
What are your strategies to start with a new DSL? Do you read their documentation, search for videos, or blog posts? Which part of the documentation do you start with? How do you navigate through the documentation? Which part of the documentation do you find most useful as a beginner, and as you get experienced? Which parts confuse you or make you overwhelmed?

\item  When do you find yourself looking into the gallery examples?  What information are you looking for? How to do a certain thing, inspirations, possibilities or general language construct? How often? In which way do you find them useful?
\item What types of tasks do you do with a gallery? Which do you find more difficult: figuring out what the design should look like or implementing it with the DSL? Which step is the documentation or gallery more helpful with?

\item Can you describe the most recent or most common process for you to create a visualization using the [DSL]? Do you start from scratch, template, examples similar to your goal? Do you integrate different examples? How did it make you more fluent with the library? What’s your leaning and usage trajectory? How long does it take? What are some decisions you made? How did you decide?
\item Which code examples are more useful to you, the ones from tutorials and library references, or the gallery examples? Or what are the different situations when you search for these code examples? 
\item What strategies do you use to find relevant examples? (For which tasks do you use each of these strategies?) Do you search for external resources? Do you browse the gallery examples? Do you search with keywords on search engines or inside the doc site? How useful are the keywords?  How did you decide what is helpful? 
\item How do you use these examples after you find them? Do you try to understand every detail before making edits? Do you look into tutorials and references specific to the example? Do you look for help from the library maintainers, the communities and Stack Overflow? Do you find them helpful?
\item What aspects are the most important when you look at an example? Environment: Being able to make changes and see the result right away (online editor) Explanations and code comments Simple and clear, easy to read
\item What aspects are the most important when you look at the gallery? Categorized, Sorted, aesthetically intriguing, Clear descriptions, Visuals that are easy to perceive, Number of available examples

\item What are the challenges for you to find relevant examples in a gallery?
\item What are the challenges for you to understand or repurpose relevant examples in a gallery?
\item Are you satisfied with the gallery of [DSL]? What’s hard or annoying about the galleries or creating visualizations in general? In an ideal world, what features or tools would you want to help you write code with [DSL]
\item Do you share examples you made with other people? For example, send pr to contribute or share blocks style posts with colleagues etc.

\item The ability to leverage carefully selected examples or the ability to recognize a good example from a collection? 

\end{itemize}

\subsection{Post Interview Survey for the Creators}
\parahead{Study introduction}
Thank you for sharing your knowledge and experience with us in the interview session. We are still learning how DSL/library developers design and manage example galleries. We would love for you to fill out this ~5-10 minute survey. 
This survey contains 6 close-ended questions and 2 open-ended questions. 

\begin{itemize}
    \item What is your occupation?
    \item Which gender identity do you most identify with? 
    \item How long have you been designing DSLs/API and libraries? Give your best estimate.
    \item How many DSLs/API and libraries have you designed? Give your best estimate. 
    \item How many galleries have you created for the DSLs/API and libraries you designed? Give your best estimate.
    \item Are we allowed to refer to you by name or not?
    \item Do you have additional feedback for the interview?
    \item Is there anyone you know would be a good fit for our interviewee?
\end{itemize}

\end{document}